\documentclass[useAMS,usenatbib,usefloat]{mn2e}
\usepackage{float}
\usepackage{amsmath}
\usepackage{amssymb}
\usepackage{rotating}
\usepackage{tabularx}
\usepackage{ gensymb }

\usepackage{pdflscape} 
\usepackage[font=footnotesize,labelfont=bf]{caption}

\newcommand{\ugr}{$(u-g, g-r)$ }
\newcommand{\ltf}{\log(\rm T_{\rm eff})}

\begin{document}

\title[\bf OB stars in Carina]{\bf The deep OB star population in Carina from the VST Photometric H$\alpha$ Survey (VPHAS+)}
\author[M. Mohr-Smith et al]{
{\parbox{\textwidth}{M. Mohr-Smith$^1$,
J. E. Drew$^1$\thanks{E-mail: j.drew@herts.ac.uk}, R. Napiwotzki$^1$, S. Sim\'on-D\'iaz$^{2,3}$, N. J. Wright$^{1,7}$,\\G. Barentsen$^{1,4}$, J. Eisl\"offel$^5$, H. J. Farnhill$^1$, R. Greimel$^6$, M. Mongui\'o$^1$, V. Kalari$^8$, Q. A. Parker$^9$, J. S. Vink$^{10}$}
}\\ \\
$^1$Centre for Astrophysics Research, Science and Technology Research Institute, University of Hertfordshire, Hatfield AL10 9AB, UK \\
$^2$Instituto de Astrof\'isica de Canarias, E-38200 La Laguna, Tenerife, Spain\\
$^3$Departamento de Astrof\'isica, Universidad de La Laguna, E-38205 La Laguna, Tenerife, Spain\\
$^4$NASA Ames Research Center, Moffatt Field, Mountain View, CA 94043, USA\\
$^5$Th\"uringer Landessternwarte Tautenburg, Sternwarte 5, D-07778 Tautenburg, Germany\\
$^6$Institute of Physics, Department for Geophysics, Astrophysics \& Meteorology, NAWI Graz, Universit\"atsplatz 5, 8010 Graz, Austria\\
$^7$Astrophysics Group, Keele University, Keele, ST5 5BG, UK\\
$^8$Departamento de Astronom\'ia, Universidad de Chile, Camino El Observatorio 1515, Las Condes, Santiago, Casilla 36-D, Chile\\
$^{9}$Department of Physics, The University of Hong Kong, Pokfulam Road, Hong Kong, China\\
$^{10}$Armagh Observatory, College Hill, Armagh, BT61 9DG, Northern Ireland 
}

\maketitle

\begin{abstract}

Massive OB stars are critical to the ecology of galaxies, and yet our knowledge of OB stars in the Milky Way, fainter than $V \sim 12$, remains patchy.  Data from the VST Photometric H$\alpha$ Survey (VPHAS+) permit the construction of the first deep catalogues of blue excess-selected OB stars, without neglecting the stellar field.  A total of 14900 candidates with 2MASS cross-matches are blue-selected from a 42 square-degree region in the Galactic Plane, capturing the Carina Arm over the Galactic longitude range $282^{\circ} \lesssim \ell \lesssim 293^{\circ}$. Spectral energy distribution fitting is performed on these candidates' combined VPHAS+ $u,g,r,i$ and 2MASS $J,H,K$ magnitudes. This delivers: effective temperature constraints, statistically separating O from early-B stars; high-quality extinction parameters, $A_0$ and $R_V$ (random errors typically $< 0.1$).  The high-confidence O-B2 candidates number 5915 and a further 5170 fit to later B spectral type.  Spectroscopy of 276 of the former confirms 97\% of them. The fraction of emission line stars among all candidate B stars is 7--8\% . Greyer ($R_V > 3.5$) extinction laws are ubiquitous in the region, over the distance range 2.5--3 kpc to $\sim$10~kpc.  Near prominent massive clusters, $R_V$ tends to rise, with particularly large and chaotic excursions to $R_V \sim 5$ seen in the Carina Nebula.  The data reveal a hitherto unnoticed association of 108 O-B2 stars around the O5If$+$ star LSS 2063 ($\ell = 289.77^{\circ}$, $b = -1.22^{\circ}$).  Treating the OB star scale-height as a constant within the thin disk, we find an orderly mean relation between extinction ($A_0$) and distance in the Galactic longitude range, $287.6^{\circ} < \ell < 293.5^{\circ}$, and infer the subtle onset of thin-disk warping.  A halo around NGC~3603, roughly a degree in diameter, of $\sim$500 O-B2 stars with $4 < A_0 (\rm{mag}) < 7$ is noted.

\end{abstract}

\begin{keywords}
stars: early-type, (Galaxy:)
open clusters and associations, (ISM:) dust, extinction, Galaxy: structure, surveys
\end{keywords}

\clearpage

\section{Introduction}

Massive OB stars are critically important objects in shaping the evolution of galactic environments.  But because of their relative rarity due to the initial mass function (IMF) and their short lives, they have proved to be difficult to nail down in terms of their evolution before and after the main sequence: their formation is still subject to debate \citep{Tanetal2014}  and we cannot say with any certainty what exact fate awaits which type of massive object \citep{Langer2012}.  It has long been recognised that the UV radiation and supernovae they produce are huge factors in shaping the interstellar medium (ISM). 

As things stand, the best-studied resolved massive star population can be claimed to be located in the Magellanic Clouds, thanks to the VLT-FLAMES Tarantula Survey \citep[][and subsequent papers]{Evansetal2011} that has produced a coherent spectroscopic dataset on over 800 massive stars.  This has captured well the properties of a population of massive stars at reduced metallicity ([Fe/H] $\simeq -0.3$). In contrast, the impact of varying levels of significant extinction has rendered the higher metallicity disk of our own galaxy more of a challenge.  Existing catalogues of the O and early-B stars are confined to objects brighter than $\sim$12th magnitude \citep[complete only to $\sim2$~kpc, see e.g.][]{Garmanyetal1982} and are heterogeneous \citep[e.g.][]{Reed2003}.  An effort is under way to collect complete spectroscopy of the known bright Galactic O-star population \citep[GOSSS by][and subsequent papers]{maiz-apellaniz2011} but, limited by the source material it presently rests on, only a few tens of objects have yet been covered over the range $12 < B < 16$ \citep{gosc2016}.   So far the known fainter reddened OB stars are mostly located in massive clusters, that are more easily noticed against the Galactic background.

There is thus a gap in our knowledge of the wider field population: this presents a problem to the debate over the significance of the runaway phenomenon \citep{Zwartetal2010} and how this should be tensioned against the possibility of forming massive stars outside densely clustered environments \citep{deWitetal2005,Bressert2012}. Furthermore, as we await the arrival of advanced astrometry from ESA's Gaia mission, it is the perfect time to trace how OB stars fit into the structure of the star-forming Galactic thin disk, reaching well beyond the solar neighbourhood.  Mapping these intrinsically luminous objects is important for another reason: they have long been valued as distant and bright back-lighters of the ISM -- much of what is known about interstellar extinction has been deduced from fitting the SEDs of OB stars \citep[e.g.][]{Cardelli1989, fitzpatrickandmassa2007, MaizA2014}.   A new deep census of OB stars has much to offer. 

\begin{figure*}
\centering
\includegraphics[width=\textwidth]{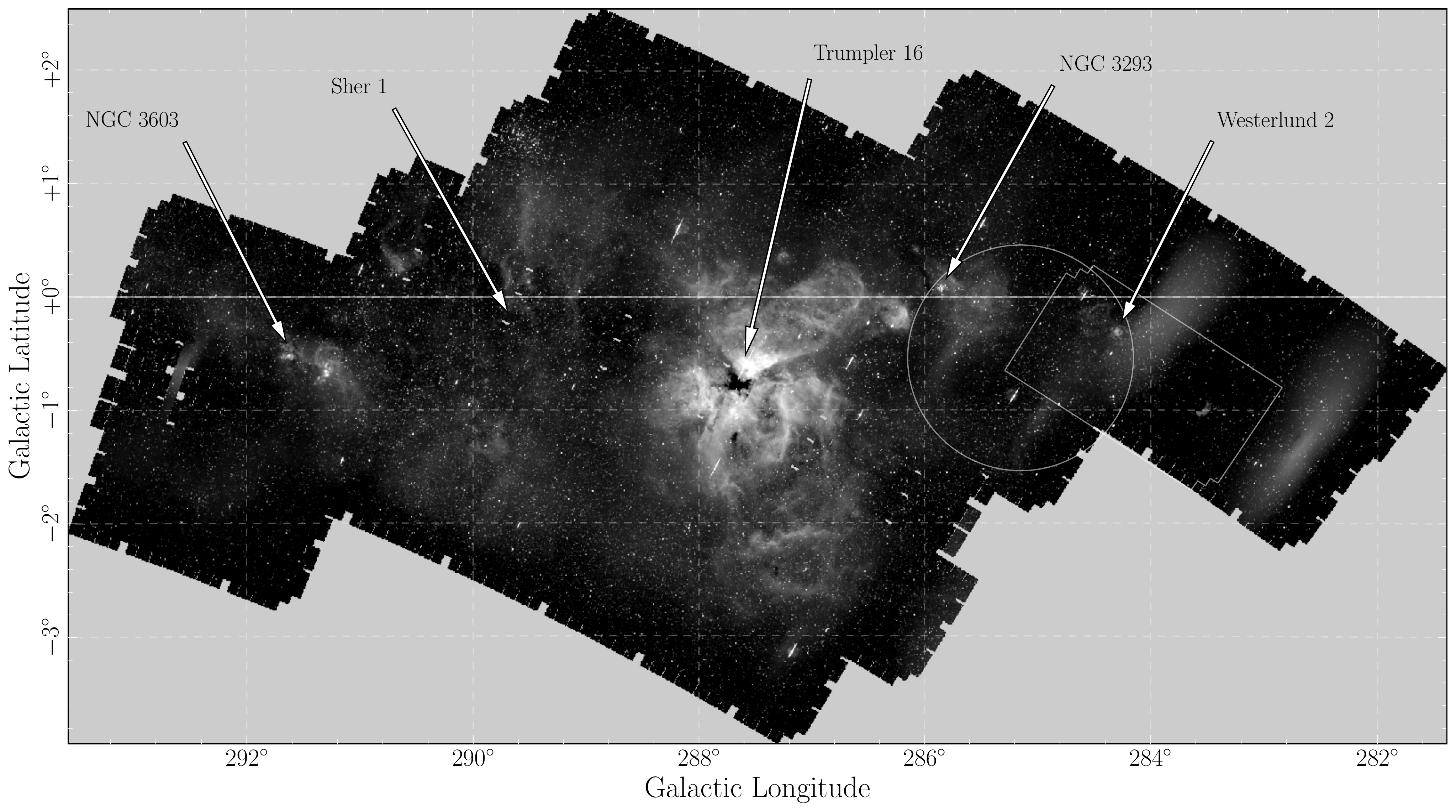}
\caption[VPHAS+ H$\alpha$ image showing the footprint of the 42 square degrees in Carina that are studied here.]{VPHAS+ H$\alpha$ image showing the footprint of the 42 square degrees in Carina that are studied here. The positions of major OB star clusters are labelled. The white rectangle shows the area studied in \protect\citep{mohr-smith2015} and the circle shows the area studied with spectroscopy in Section \protect\ref{sec:spectra}. The two large bright smudges in the far right, and that on the far left of the footprint are due to scattered light in the constituent images that are not pipeline corrected for illumination effects (which are included in the photometric catalogue generation). }
\label{fig:carinaregion}
\end{figure*}

The cheapest method of astronomical census is provided by photometry. At the time \cite{johnsonandmorgan1953} introduced the $UBV$ photometric system they also described the concept of exploiting the ($U-B$,$B-V$) colour-colour (CC) diagram to construct reddening-free parameters labelling reddening lines for distinct stellar spectral sub-types.  This introduced a method of analysing stellar photometry that can now be applied with greatly increased power to very large, homogeneous data sets in this new era of survey photometry.  It works especially well as a tool to select hot massive OB stars: the aim of this paper is to apply a customised version of it to newly-minted data from the VST\footnote{The VST is the VLT Survey Telescope, and the VLT is the Very Large Telescope, a set of 4 8-m telescopes operated by the European Southern Observatory} Photometric H$\alpha$ Survey of the Southern Galactic Plane and Bulge \citep[VPHAS+,][]{drewetal2014}.  The target region for this work, intended both as an up-scaling of our already validated method and as an exploration of a part of the Milky Way particularly rich in massive OB stars, is the Carina region of the Galactic plane.  

In a previous paper \citep{mohr-smith2015}, we presented a first blue selection from VPHAS+ data that focused on a region spanning just 2 square degrees, around the massive O-star-rich cluster, Westerlund~2.  This adapted the method of reddening-free parameters to the Sloan filter system in use at the VST and performed a validation test on a catalogue of objects, that had been selected from the ($u-g$,$g-r$) diagram, and further winnowed through fitting $u$,$g$,$r$,$i$ and published $J$,$H$,$K$ magnitudes to progressively-reddened synthetic stellar photometry.  The haul from this was a list of 489 objects with spectral types of B2 or earlier -- most of them were new discoveries, representing an order of magnitude increase on the total known from previous catalogues.

It became clear in this first pilot that the characterisation of extinction is particularly high in quality: the typical precisions achieved are 0.1 or better in each of $A_0$, the monochromatic extinction in magnitudes at 4595~\AA , and $R_V$, the ratio of total to selective extinction.  The new, expanded catalogue of candidate OB stars included both discoveries within Westerlund~2 itself and a handful of similarly reddened O stars scattered around it at offsets of between 10 and 40 arcminutes.  The latter may have been ejected from the cluster or they may be evidence of a wider star-forming event within which Westerlund 2 is the most prominent feature.  To settle this, spectroscopic follow up is required.  The new work here expands the sky area from 2 square degrees up to 42 and the number of high-confidence candidate O to B2 stars from 489 to 5915.

The total footprint spanning the Carina Nebula and environs is shown in Fig.~\ref{fig:carinaregion}: it runs from close to the arm's tangent direction, at $\ell \sim 282^{\circ}$ \citep{Dame2007, Vallee2014}, through Westerlund 2, and on through the Carina Nebula, to beyond the brilliant massive open cluster, NGC 3603, ending at $\ell \simeq 293^{\circ}$. The irregular footprint outline, with reduced latitude coverage at lower longitudes, reflects the processed data available at the outset of this work. The longitude range examined is much the same as that investigated by Graham (1970) who used Walraven photometry to map 436 bright ($8.5 \lesssim V \lesssim < 11.5$) OB stars.  Indeed, the work we present amounts to an update, almost 50 years on, that reaches from 10 to more than 1000 times fainter.

On the map of Galactic spiral arm structure presented by \cite{Russeil2003} the Carina Arm is the single most compelling and coherent feature.  In CO maps \citep{Grabelsky1987, Dame2001} the arm is traceable as a somewhat broken ridge of emission that has a near side about 2~kpc away, running from higher Galactic longitude to the tangent point 4--6 kpc distant \citep[see the discussion by][]{Dame2007}: it connects to the far side beyond the tangent as it carries on unwinding out beyond the Solar Circle: \cite{Grabelsky1987} suggested heliocentric distances of the order 10--12~kpc for the arm at $\ell \sim 293^{\circ}$.  The famous Carina Nebula is generally regarded as being embedded in the near portion of the arm, 2.5--3 kpc away \citep{tapiaetal2003, huretal2012, kumaretal2014}, while NGC~3603 is associated with the far arm, $\sim$7 kpc away \citep{sungandbessell2004}.  It aids visualisation of the arm to notice that 11 degrees of longitude corresponds to roughly 1.1 kpc at $\sim$5 kpc distance -- not much of a bend over the potential heliocentric distance range to $\sim$10~kpc.  Nevertheless, a little curvature, turbulent motions and recent star formation create complexity in the dust distribution that demands high density sightline sampling.  

This paper is organised as follows: the first major section, section 2, presents the photometry and how it is exploited to select and characterise the optically-fainter Carina OB-star population; the results of the selection are presented in section 3 along with exercises aiming to distinguish the minority of non main sequence objects included in it; the quality of the selection is then examined more closely in section 4 by analysing follow-up spectroscopy of a subset of candidates located within a single 2 degree field as marked on Fig.~\ref{fig:carinaregion}. The outcome is that almost all the high confidence candidates are confirmed. Finally in section 5, the new catalogue of 5915 candidate OB stars is put to use in an appraisal of the trends in extinction ($A_0$) and extinction law ($R_V$)across the studied region.  We then experiment with adopting the OB-star scale height within the thin disk as a standard ruler in order to set a scale to the rise in extinction with heliocentric distance.  This works over the longitude range $287.6 ^{\circ}<\ell<293.5 ^{\circ}$ and leads in turn to measures of the onset of the warp of the Galactic thin disk.  The paper ends with some discussion and the main conclusions in section 6.

\section{Method: the optical photometry; OB-star selection and SED fitting}

\subsection{VPHAS+ optical photometry}

The optical photometry we exploit is from the VPHAS+ survey \citep[described in full by][]{drewetal2014}.  This survey of the southern Galactic Plane began at the end of 2011 and, at the time of writing, continues to execute on the VST at the Paranal site in Chile.  The VST camera, OmegaCam, delivers $1\times1$ square-degree images of the sky at a typical seeing of 0.8--0.9 arcsec in the Sloan $gri$ and narrowband H$\alpha$ filters, rising to 1.0--1.1 arcsec in Sloan $u$.  In all bands, the limiting (Vega) magnitude is typically greater than 20th, reaching as faint as 22nd in $g$.   All data from the survey are pipeline-reduced to source lists, calibrated against nightly standards, by the Cambridge Astronomical Survey Unit (CASU).  These are used in this study in the form of band-merged $ugr$ and $riH\alpha$ contemporaneous catalogues. We note that the blue and the red filters are usually observed months apart -- the repeated $r$ band observations are in place to serve as basic checks on variability.  The raw materials for this study, are the collated source catalogues for a set of 42 survey fields spanning the chosen Carina Galactic plane region (Fig.~\ref{fig:carinaregion}).

A new feature of expanding to the larger set of fields is that it is even more important to impose a uniform photometric calibration across all the data.  To achieve this, we have made comparisons on a field by field basis with $g$, $r$ and $i$-band stellar photometry from The AAVSO Photometric All-Sky Survey (APASS) and have then used the main stellar locus in the \ugr  colour-colour (CC) diagram to determine the $u$ band shifts required.  The $H\alpha$ band data are tied to the $r$ band calibration simply by requiring the H$\alpha$ photometric zeropoint to be offset from the $r$ zero-point by 3.08.  The details of this alignment process will be presented in a forthcoming presentation of VPHAS+ data products (Drew et al, in prep). The mean photometric adjustments made to the broadband data, as a result of these comparisons, are given in Table~\ref{table:cumhistcalib}.  The sense of all the tabulated shifts is such that the revised magnitude scale is brighter than the unrevised.  In this paper, we express all magnitudes in the Vega system.


We have exploited the field overlaps provided by the tiling pattern of VPHAS+ \citep[see][]{drewetal2014} in order to check the final uniformity of the recalibration. Within the overlap regions, it is possible for a star to have up to four independently recalibrated detections in each band.  Using these, we can evaluate the convergence to a common scale by comparing the standard deviation of multiple measurements before and after the calibration adjustments.  These are set out, for the broad bands, in Table \ref{table:cumhistcalib}: there is noticeable
improvement in all of them and especially in $g$.

\begin{table}
\caption{Mean zero-point calibration shifts in each broad band, and the standard deviations of repeated detections in overlaps, before and after adjustment to bring all fields onto a common photometric scale.}
\centering
\begin{tabular}{cccc}
\hline
Band & mean zero-point shift & $\sigma_m$ before & $\sigma_m$ after \\
\hline
$u$ & 0.38 & 0.029 & 0.027 \\
$g$ & 0.04 & 0.030 & 0.015 \\
$r$, H$\alpha$ & 0.09 & 0.027 & 0.019 \\
$i$ & 0.07 & 0.028 & 0.024 \\
\hline
\end{tabular}
\label{table:cumhistcalib}
\end{table}

\begin{figure*}
\centering
\includegraphics[width=0.98\textwidth]{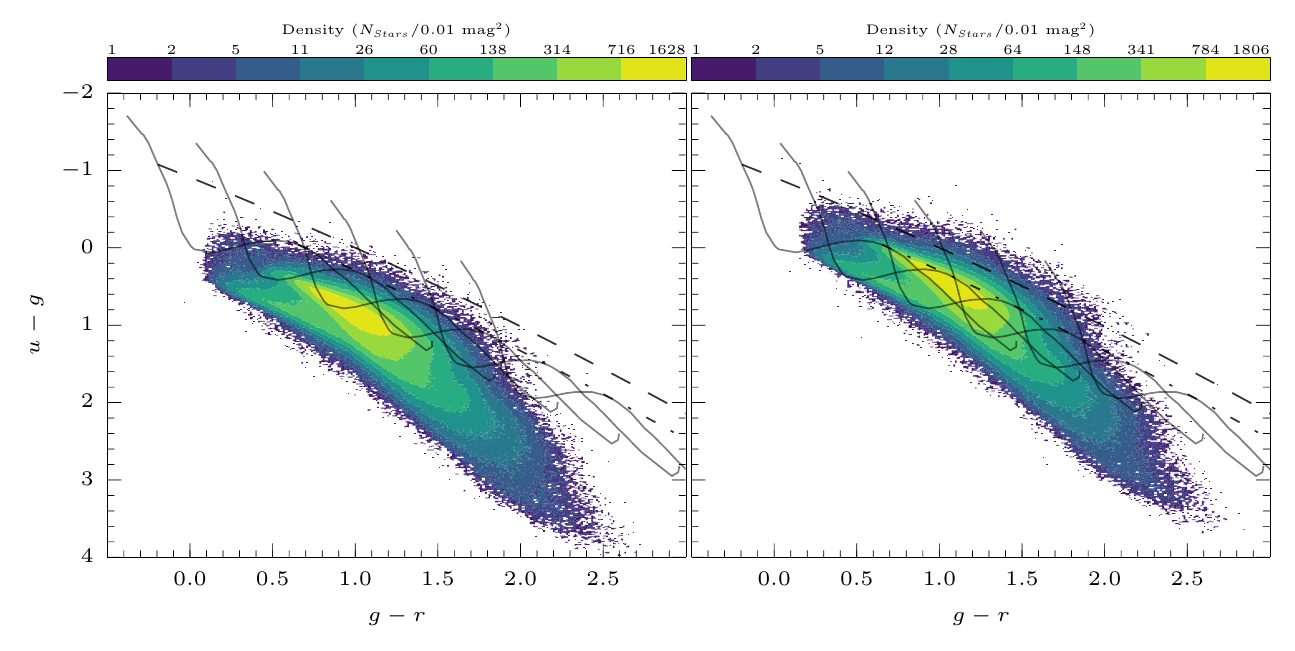}
\includegraphics[width=\textwidth]{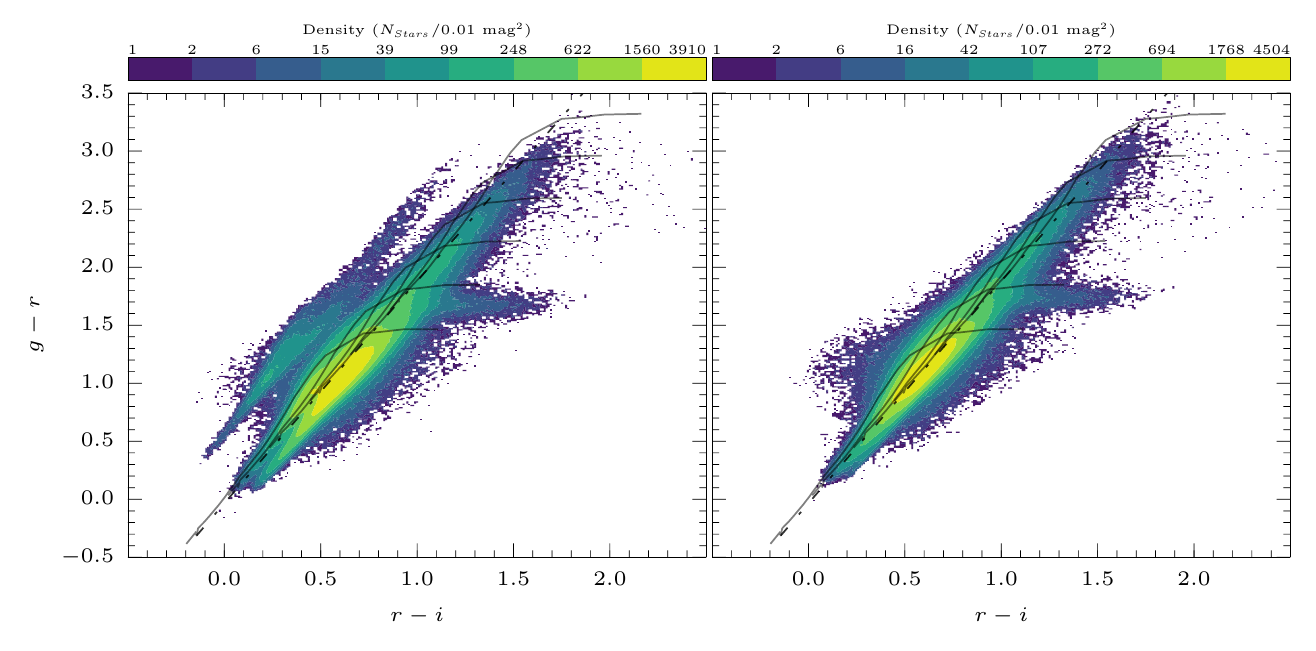}
\caption{Colour-colour density plots showing the position of the main stellar locus with respect to the synthetic main sequence ($A_0 = 0 - 10$ magnitudes in steps of 2) before (left) and after (right) re-calibration. The dashed-dot line shows the G0V reddening vector while the dashed line shows the B3V reddening vector. The left lower panel clearly shows the presence of poorly calibrated exposures that are brought into line by our calibration corrections. The main correction of the upper left panel is an upward shift in $u-g$ linked to a systematic brightening of the $u$ band data (see table~\ref{table:cumhistcalib}).}
\label{fig:calib*}
\end{figure*}

As final confirmation of the improved uniformity of the data, after correction,
we show in Fig. \ref{fig:calib*} before and after source density plots of
the ($u-g, g-r$) and ($r-i$, $g-r$) CC diagrams for the region studied here. We can
see that the diagrams have much tighter distributions with less scatter after
correcting the individual field calibrations. The main-stellar locus (highest
density of sources) is also placed in the correct position with respect to
the synthetic tracks.  The data in Table~\ref{table:cumhistcalib}, alongside Fig.~\ref{fig:calib*}, show that the main consequence of the $u$
re-calibration is to bring this band into better (astrophysical) alignment
with the longer-wavelength bands, rather than to reduce scatter.  For the
successful selection of candidate OB stars, this is critical.

\subsection{OB-star selection and SED fitting}

The methods for the selection of candidate OB stars and spectral energy distribution (SED) fitting are the same as those used and described in full by \cite{mohr-smith2015}. Below, we summarise the main features.

\subsubsection{Photometric Selection and Cross Matching}

First, a \ugr CC diagram is constructed that includes all catalogued objects with $g \leq 20$, a $u$ detection and random photometric magnitude errors below 0.1 in $g$, $r$ and $i$.  OB stars in this diagram are offset above the main stellar locus and can be selected using a method related to the well-established Q method of \cite{jandm1953}. We select targets blue-ward of the B3V reddening vector, i.e. above it, as drawn in the \ugr CC diagram in Fig.~\ref{fig:calib*}  \citep[see also Fig. 3 in][]{mohr-smith2015}. This aims at picking out all objects earlier than B3 (our target group). The number of objects in this initial selection was 37971.

All candidates are then cross-matched against the Two Micron All Sky Survey (2MASS), the only available nIR survey covering the whole region. The maximum cross match distance was set to 1 arsec in forming optical near-infrared (OnIR) SEDs. As long as the nIR magnitudes of the the initially selected objects are within the 2MASS range, the crossmatching is very successful since the VPHAS+ survey astrometry is calculated with reference to 2MASS. The 90$^{\rm th}$ percentile cross match separation is 0.3 arcseconds and the median is 0.073 arcseconds.  Only objects with catalogued magnitudes in all seven bands ($u, g, r, i, J, H, Ks$) are retained for input into the SED fitting procedure.

\subsubsection{SED fitting}

We compare the empirical SEDs to a parametrized model in a simple Bayesian scheme using Markov Chain Monte Carlo (MCMC) sampling. The model consists of four parameters $\ltf$, $A_0$, $R_V$ and $\mu$, the distance modulus. The intrinsic SEDs are taken from the Padova isochrone database \citep[CMD v2.2 \footnote{http://stev.oapd.inaf.it/cgi-bin/cmd};][]{Bressanetal2012, Bertellietal1994}. As the OnIR colours of OB stars do not vary significantly with luminosity class or metallicity \citep{M&P2006}, only main-sequence, solar-metallicity models have been considered ($\log(g) \sim 4$, $Z = 0.019$).  The models are then reddened using a set of \cite{fitzpatrickandmassa2007} reddening laws and shifted according to distance modulus. As the apparent OnIR SEDs of OB stars change only subtly as a function of effective temperature \citep{M&P2006} the shape of their SEDs is controlled mainly by extinction. This means that $\ltf$ is relatively weakly constrained, whilst permitting reasonable O/B separation (see Section \ref{sec:spectra}).  The extinction parameters, on the other hand, are well constrained. As there is no luminosity class discrimination, $\mu$ acts primarily as a normalization factor and is highly correlated with effective temperature \citep[see Fig. 6 of][]{mohr-smith2015}.

The likelihood function used in the Bayesian scheme is a multivariate Gaussian that embeds the assumption that the uncertainties on all measurements/parameters are uncorrelated\footnote{This is not strictly true for $\mu$ and is further reason to treat this parameter with caution}. This reduces to the familiar sum for $\chi^2$:
\begin{equation}
\label{eqn:chi}
P \left( SED_{obs} \mid \theta \right) \propto \exp\left(-\frac{1}{2}\sum\limits_{i}^{n} \frac{(m(obs)_i - m(mod)_i)^2}{\sigma_i^2}\right)
\end{equation}
where $P \left( SED_{obs} \mid \theta \right)$ is the probability of obtaining the observed SED ($SED_{obs}$) given a set of parameters $\theta$, and where $m(obs)_i$ and $m(mod)_i$ are the observed and model magnitudes in each band $i$.  In this equation only, $i$ is just a summation index

We have adopted uniform priors on each parameter which are as follows:
\begin{multline}
\label{eqn:params}
P(\theta) = \begin{cases} 1 \quad \text{if} \,\,\begin{cases} 4.2 \,\, \leq \,\, \ltf \,\, \leq \,\, 4.7 \\
0 \,\, \leq \,\, A_0 \,\, \leq \,\, 15 \\
2.1 \,\, \leq \,\, R_V \,\, \leq 5.1 \\
0 \,\, \leq \,\, \mu \,\, \leq \,\, 20
 \end{cases}\\
 0 \quad \text{else}
 \end{cases}
\end{multline}
The lower limit on $\ltf$ is chosen to be below the temperature of a B3V star \citep[cf. table 3 in][]{nieva2013} and the upper limit is the maximum temperature of a MS star in the Padova isochrones. The latter corresponds to earlier than O3V \cite[see e.g.][]{Martinsetal2005}. The upper limit on $A_0$ is chosen to comfortably enclose the likely extinction range among the selected OB stars, given the $g=20$ limiting magnitude, assuming a typical rate of rise in $A_0$ of $\sim 1.5$ mag kpc$^{-1}$. The limits on $R_V$ cover what we expect to find in the Galaxy \citep{fitzpatrickandmassa2007} and the upper limit on $\mu$ amounts to leaving it unbound. The MCMC fitting procedure was performed using the \textsc{python} package \textsc{emcee} \citep{Foreman-Mackeyetal2013}.

As in \cite{mohr-smith2015}, SED fits with $\chi^2 < 7.82$ are taken to be acceptable: the probability of achieving this value in the case of the null hypothesis is 0.05 for the 3 degrees of freedom involved in this fitting process. 

\section{Results of the photometric selection}
\subsection{Overview}
\label{sec:carina_intro}

Table~\ref{tbl:carina_breakdown1} shows the breakdown of the number of OB candidates according to effective temperature and fit quality. In total 14900 OB stars were selected for SED fitting. Around a quarter of the candidates where found to have poor, $\chi^2 > 7.82$, SED fits. This leaves a clear majority of object with acceptable SED fits. Just under half of these are probable B3 and later type stars with $\ltf<4.30$. So finally, there are 5915 objects in our target effective temperature range, $\ltf \geq 4.30$ with $\chi^2 < 7.82$. Of these, 905 are probable O stars with $\ltf \geq 4.477$. There are thus around five times as many B2--B0 star candidates as O star candidates -- a ratio that exceeds the IMF ratio of $\sim$3, thanks in part to the bright magnitude limit set by the survey data.

All candidates were cross matched with SIMBAD to within an acceptance radius of 1 arcsec in order to check if any objects with already known spectral type are in the photometric selection. Table~\ref{tbl:carina_breakdown2} shows a numerical breakdown, according to derived effective temperature and fit quality, of the subset of OB candidates with spectroscopically confirmed types according to SIMBAD.  These are supplemented with the spectroscopically confirmed types from our AAOmega spectroscopy to be presented in Section \ref{sec:spectra}.  The majority of objects with known spectral type from SIMBAD were found to be OB stars as expected, with some WR stars appearing in the $\chi^2 > 7.82$ groupings. There is also contamination from a handful of M giant stars and one carbon star (C*) among the poor fits. These objects have mimicked very highly reddened OB stars in the $(u-g,g-r)$ diagram but nevertheless they are easily distinguished in the selection both via SED fit quality and via $r-i$ colour as discussed in Section \ref{sec:emissioncarina}. A very positive feature of tables~\ref{tbl:carina_breakdown1} and \ref{tbl:carina_breakdown2}, in combination, is the high success rate in placing stars correctly into the O and B spectral types. Inevitably the application of a cut in $\chi^2$ will exclude some genuine O and early B stars, either randomly or because of source blending and/or the presence of circumstellar emission.

\begin{table}

\caption{Breakdown of the number of OB candidates according to effective temperature and fit quality.}

\label{tbl:carina_breakdown1}

\begin{center}

\begin{tabular}{ccc}

\hline \noalign{\smallskip}

& $\chi^2 \leq 7.82$ & $\chi^2 > 7.82$ \\ \noalign{\smallskip}

\hline \noalign{\smallskip}

O Stars: $\ltf \geq 4.477$  & 905 & 1090\\ \noalign{\smallskip}

B2 - B0: $4.30 \leq \ltf < 4.477$ & 5010 & 1039\\ \noalign{\smallskip}

Late B: $\ltf < 4.30$ & 5170 & 1686\\ \noalign{\smallskip}

\hline \noalign{\smallskip}

All $\ltf$ & 11085 & 3815\\ \noalign{\smallskip}

\hline

\end{tabular}


\caption{Breakdown of the number of selected OB candidates with spectroscopically confirmed types according to SIMBAD and those with spectroscopically confirmed types from AAOmega in Section \ref{sec:spectra}}

\label{tbl:carina_breakdown2}

\begin{tabular}{rcc}

\hline \noalign{\smallskip}

& $\chi^2 \leq 7.82$ & $\chi^2 > 7.82$ \\ \noalign{\smallskip}

\hline \noalign{\smallskip}

$\boldsymbol{\ltf \geq 4.477}$ & & \\ \noalign{\smallskip}

AAOmega Confirmed O & 18 & 0\\
        Confirmed B & 5 & 2\\
\hphantom{AAOmega 111111}Other & 3 A/F/G & 9 A/F/G \\ \noalign{\smallskip}

SIMBAD  Confirmed O & 23 & 7 \\
        Confirmed B & 3 & 4 \\
\hphantom{SIMBAD 1111111}Other & 2 WR & 3 WR \& 7M\\ \noalign{\smallskip}

\hline \noalign{\smallskip}

$\boldsymbol{4.30 \leq \ltf < 4.477}$ & & \\ \noalign{\smallskip}
AAOmega Confirmed O & 10 & 1 \\
        Confirmed B & 174 & 10 \\
\hphantom{AAOmega 111111}Other & 6 A/F/G & 2 A/F/G \\ \noalign{\smallskip}
SIMBAD  Confirmed O & 4 & 2 \\
        Confirmed B & 18 & 15 \\

\hphantom{SIMBAD 1111111}Other & 1WR & 13 WR \& 1 M \& 1 C* \\ \noalign{\smallskip}

\hline \noalign{\smallskip}

$\boldsymbol{\ltf < 4.30}$ & & \\ \noalign{\smallskip}
AAOmega Confirmed O & 0 & 0 \\
        Confirmed B & 3 & 2 \\
\hphantom{AAOmega 111111}Other & 0 & 2 A/F/G \\ \noalign{\smallskip}
SIMBAD  Confirmed O & 0 & 0 \\
        Confirmed B & 6 & 1 \\

\hphantom{SIMBAD 1111111}Other & 0 & 10 WR \& 1 M\\  \noalign{\smallskip}

\hline

\end{tabular}

\end{center}
\end{table}

\begin{figure*}
\centering
\includegraphics[width=\textwidth]{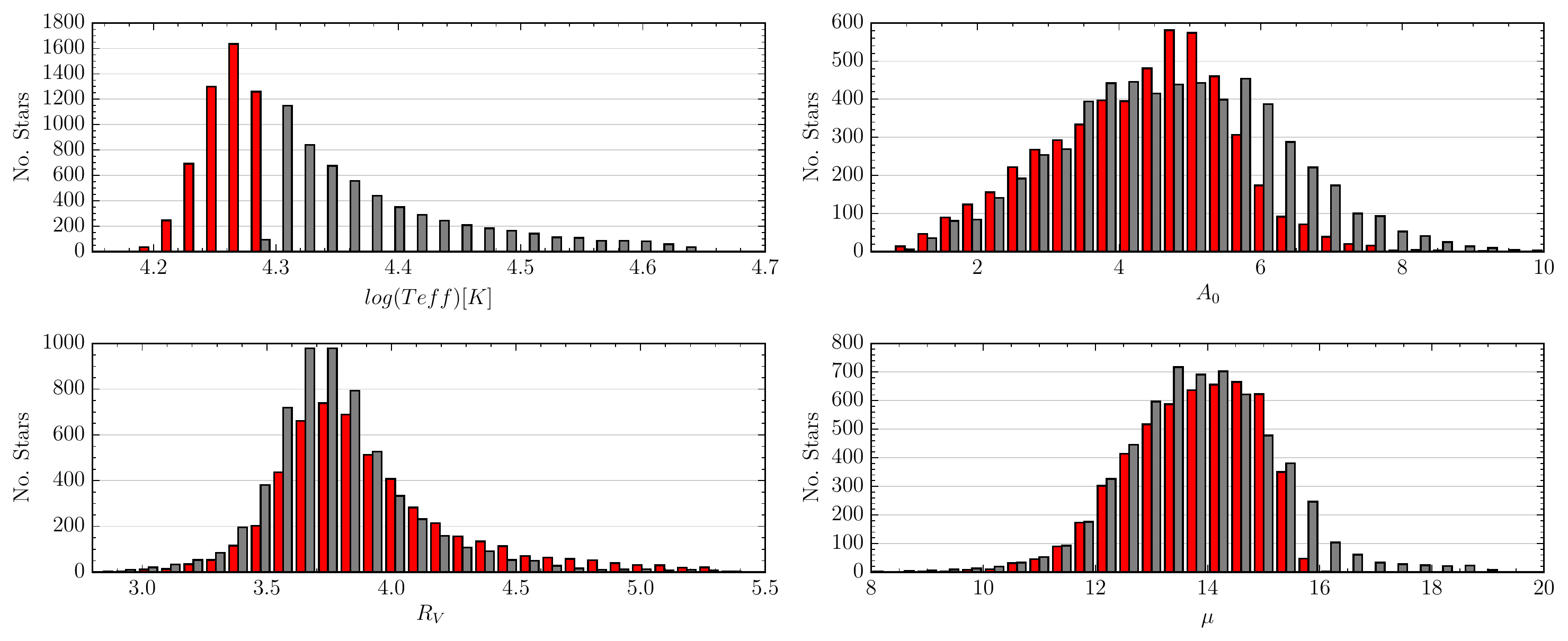}
\caption{Distribution of the best fit parameters for the selection of objects with $\chi^2<7.82$. Objects with $\ltf > 4.30$ are coloured in grey. The remaining cooler objects are coloured red. The turn over in the $\ltf$ distribution at $\sim4.26$ is a product of the initial photometric selection seeking to exclude cooler B3 and later type stars.}
\label{fig:paramhist}
\end{figure*}

\begin{figure}
\centering
\includegraphics[width=\columnwidth]{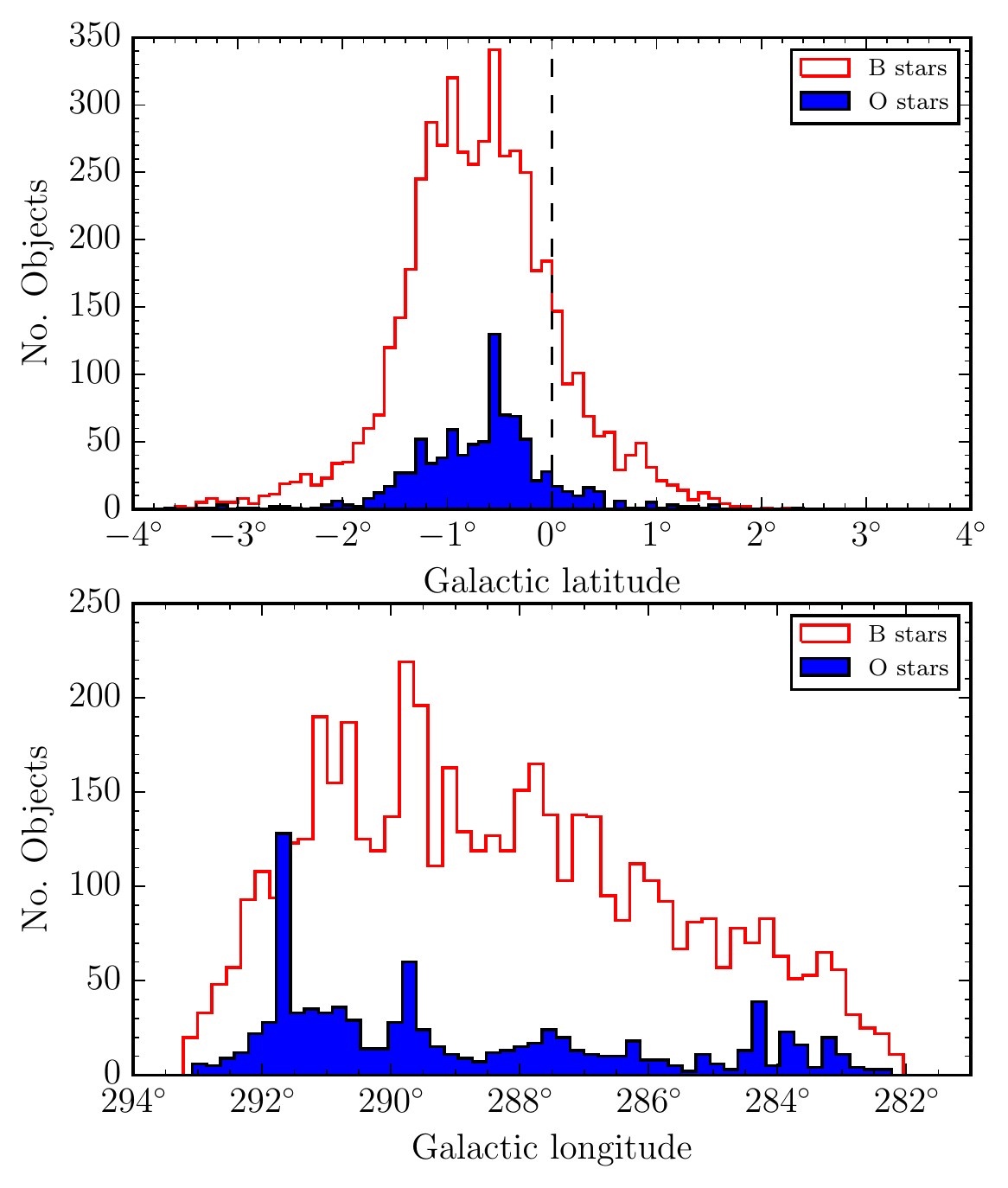}
\caption{The distribution in Galactic latitude and longitude of O and early B star candidates ($\chi^2<7.82$). The B star distribution is in red and the O star distribution is in blue (filled). The dashed line marks $b=0$. The strongest peaks in the O star distributions in both panels are associated with the NGC 3603 region. \label{fig:carina_OB}}
\end{figure}

Fig.~\ref{fig:paramhist} shows the distributions of parameters for all objects with acceptable ($\chi^2 < 7.82$) SED fits. Objects with $\ltf \geq 4.30$ that are our main focus are coloured grey. As was found for the test case by \cite{mohr-smith2015}, the entire region requires extinction laws with enhanced values of $R_V$ with a median value of 3.74. The majority of the candidates have best-fit extinctions in the range ($3 < A_0 < 6$) at likely distances of 3 - 10 kpc. The Galactic longitude and latitude distributions of the accepted OB stars are shown in Fig.~\ref{fig:carina_OB}. There is evidently a broad peak in the latitude distribution between the Galactic equator and $b \simeq -1.5^{\circ}$. The logitude distribution, on the other hand, shows a gradual rise with increasing separation from the tangent direction. The sharp O star peak in the latter, sitting close to $\ell = 292^{\circ}$ is associated with NGC 3603.

\begin{figure}
\centering
\includegraphics[width=\columnwidth]{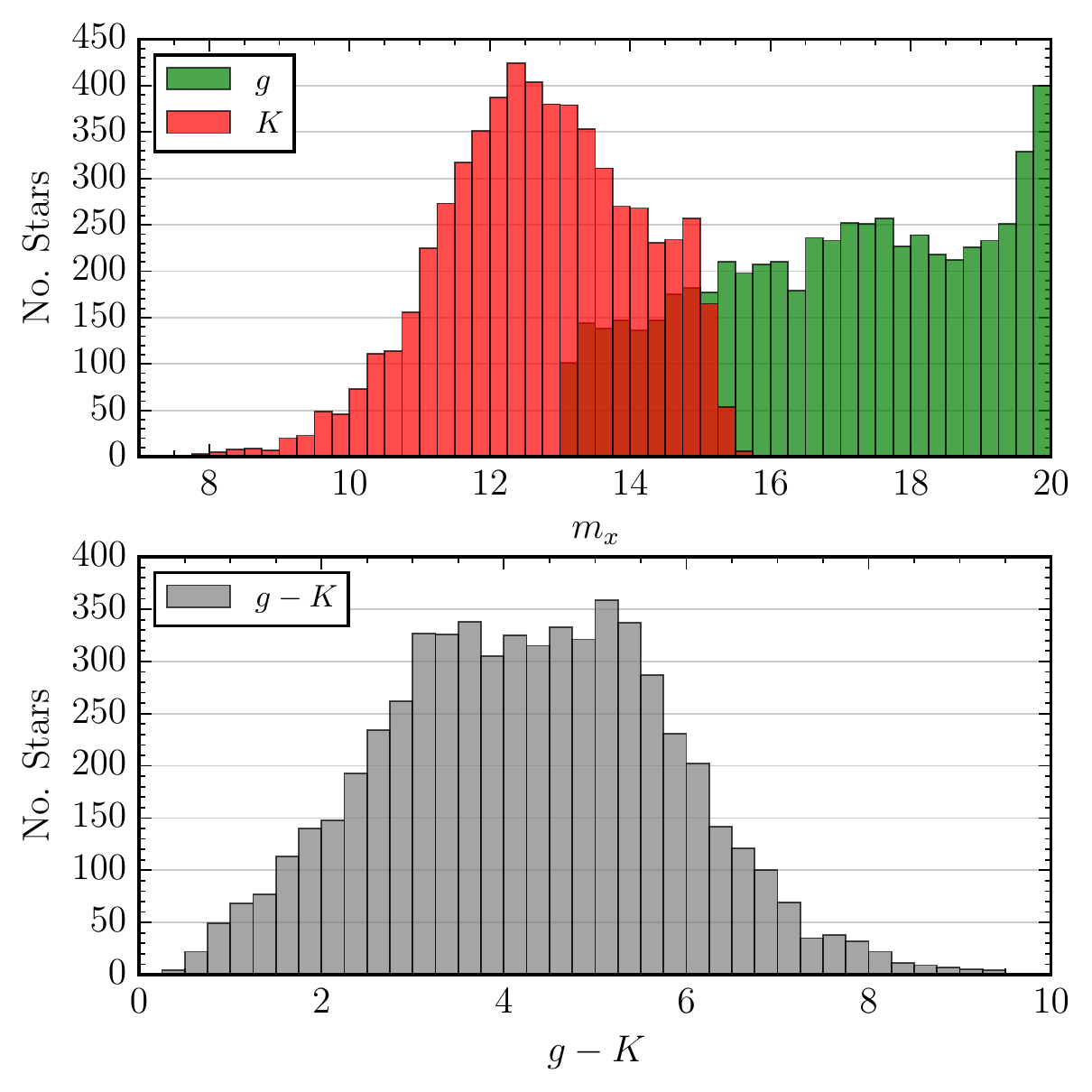}
\caption{Top panel: $g$ and $K$ band magnitude distribution of $\chi^2<7.82$ $\ltf > 4.295$ objects. Bottom panel: $g-K$ distribution.}
\label{fig:maghist}
\end{figure}

The restrictions on the magnitude range of the photometry sourced from VPHAS+, in combination with photometry from 2MASS, inevitably carry through to some constraint on the properties of the candidates uncovered by the selection and SED-fitting process.  This is illustrated in Fig. \ref{fig:maghist}, showing the distribution of $g$ band magnitudes and $g-K$ colours of all candidates with acceptable B3 or earlier-type SED fits. The fact that the magnitude ranges of the VPHAS+ $g$ and 2MASS $K$ band only overlap between $13\lesssim m \lesssim 15$ prevents us from detecting a greater number of intrinsically fainter and less reddened hot stars with lower $g-K$. As we are targeting intrinsically-bright massive OB stars, rather than sdO and sdB stars, this is not detrimental to our intended goal.

A notable feature of the $g$ distribution is the sharp upturn in the number of candidates as the $g=20$ faint limit is approached. This effect can be attributed to the growing number of sub-luminous objects at these faint magnitudes as discussed in detail in Section \ref{sec:sublum}. At $g > 19$, we are seeing the mixing of two separate distributions; one for the massive OB stars and another for the sdO and sdB stars.  If the faint $g$-band limit on the selection had been raised from $20$ to $21$st magnitude, it is likely that some more highly-reddened OB stars would have been found, along with a higher proportion of sub-dwarfs.  Nevertheless, the main motive behind setting the selection limit at $g = 20$ was to avoid growing incompleteness due to the increasing incidence at high extinction of $u$ non-detections (the typical VPHAS+ $5\sigma$ faint limit is between 21 and 21.5 in $g$, in the Vega system).

A database containing information on all 14900 candidate objects is provided as supplementary material suitable for sorting by e.g. $\chi^2$. An explanation of the database content is set out in the Appendix \ref{tbl:photom_params}. The adopted naming convention is of the form `VPHAS\_OB1\_NNNNN', where `NNNNN' is the identification number for individual objects ordered by Galactic Longitude.

\subsection{The identification of non-MS OB stars and emission line objects} 

The SED fitting performed on our blue photometric selection could not distinguish luminosity class, nor did it make practical sense to take into consideration the possibility of line emission.  So now we use evidence from the distributions of SED fit parameters to relax the MS assumption of the SED fits, and then bring in narrow-band H$\alpha$ photometry to distinguish the emission line stars.

\subsubsection{Sub- and Over-luminous Stars}

\label{sec:sublum}

\begin{figure*}
\centering
\includegraphics[width=\textwidth]{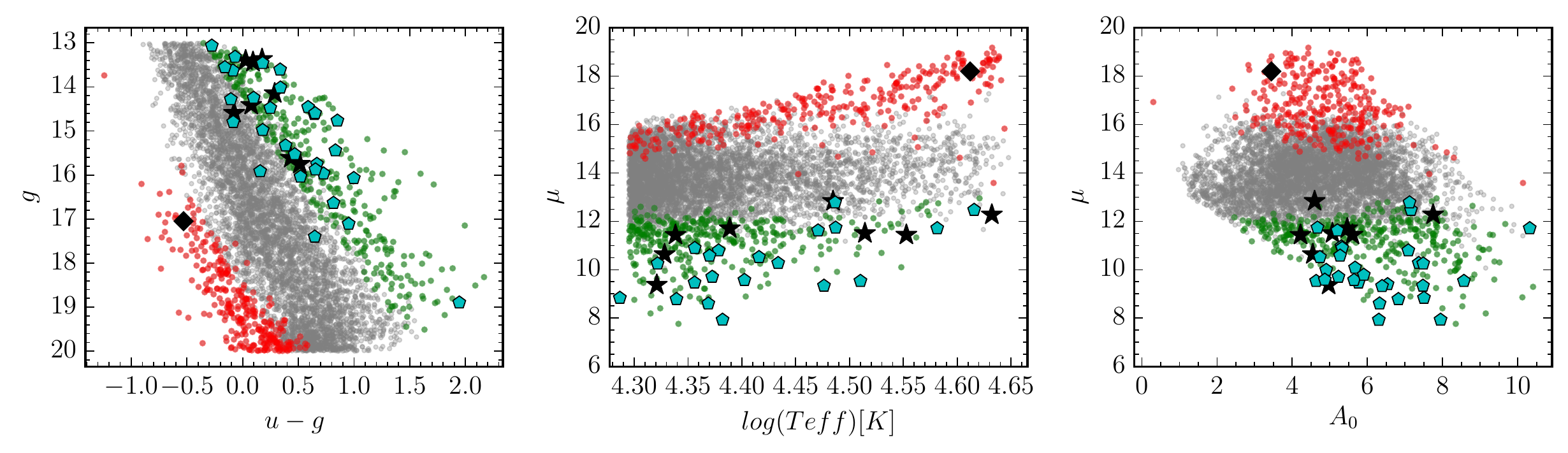}
\caption[Objects coloured in red are likely to be sub-luminous compared to the rest of the population. Objects coloured in green are probable over-luminous objects.]{Objects coloured in red are likely to be sub-luminous compared to the rest of the population. Objects coloured in green are probable over-luminous objects. The black diamond is the candidate sdO object with a spectrum presented in Sec.~\ref{sec:spectra}. The cyan pentagons pick out known WR stars and the black star symbols represent known/confirmed luminosity class I and II OB stars.}
\label{fig:sublum}
\end{figure*}

Due to the large numbers of objects included in this selection it becomes more difficult to separate the sub-luminous and over-luminous stars in the same manner as in \cite{mohr-smith2015}. This is because the different luminosity groups drawn from a much wider sky area start to mix in the $\mu~vs~A_0$ and $\mu~vs~\log(T_{\rm eff})$ diagrams used in the selection. Here, a different technique is necessary.

We can separate sub-luminous and over-luminous stars from our MS OB star targets primarily by using the colour-magnitude diagram (CMD). In the ($g$, $u-g$) CMD the sub-luminous and over-luminous stars will, for a given temperature and distance, form separate sequences below and above the main sequence. As we are looking at a range of distances these sequences become somewhat blended. Hence we aim to make a crude cut that errs on the conservative side.

First, we take all objects with acceptable $\chi^2<7.82$ fits and bin the data into deciles in effective temperature. For each decile, we then take the lowest effective temperature and the reddening vector corresponding to it, and place that vector at the $95^{th}$ percentile in distance modulus (based on all temperatures in the decile) on the ($g$, $u-g$) CMD. All stars fainter than this plausible lower-limit line are selected as sub-luminous (see Fig. \ref{fig:sublum}): they are statistically fainter than the coolest and most distant MS stars in their temperature bin. We also take the highest temperature in each bin and place it at the $5^{th}$ percentile in distance modulus on the ($g$, $u-g$) CMD. All stars brighter than this line are selected as over-luminous: they are statistically brighter than the hottest and closest MS stars in their temperature bin. 

Figure \ref{fig:sublum} shows the CMD, ($\ltf$ vs $\mu$) and ($A_0$ vs $\mu$) with likely MS stars in grey and the selected sub-luminous and over-luminous stars in red and green respectively. These distributions are compared with the positioning of examples confirmed as either sub- or over-luminous relative to the MS assumption.  In all three diagrams the known evolved OB stars reassuringly tend to co-locate in the diagram with the over-luminous candidates, whilst the one confirmed lower luminosity object also falls in the right zone.

The impact of the MS assumption in SED fits to the under-luminous objects is to make them appear as very distant objects with relatively low reddening. The absolute magnitudes of sdO stars range from $M_V=3-6$ \citep{starkandwade2003} -- some $\sim6$\,mag fainter than MS OB stars for a given temperature.  If this adjustment is made, their inferred distance moduli reduce to $\sim10-12$ from $\sim16-18$. Viewed in these terms, these objects would most likely be significantly reddened sub-dwarfs to the foreground of the main massive OB population. This is tenable for the less reddened sub-luminous candidates, but especially where $A_0$ is 4 or more (a typical value for the MS OB stars) it becomes more likely that any downward correction of distance modulus needs to be more modest -- which in turn hints at classification as e.g. post-AGB, rather than lower mass blue horizontaal branch.  We remind there is a bias to select more highly reddened subdwarf stars due to the 2MASS faint limit (see Section \ref{sec:carina_intro}).

Conversely, in Fig.~\ref{fig:sublum}, the over-luminous objects appear erroneously as close-by with higher-than-average reddening. If these objects are, for example, blue supergiants their absolute visual magnitudes would be around $\sim-6.5$ \citep{crowtheretal2006}. If these are adopted in place of MS values, their derived distance moduli, $\mu$, would rise from $\sim10$ to $\sim14$, placing them in the midst of the general near-MS massive OB population.  From our own spectroscopy (Sec.~\ref{sec:spectra}) and also from the literature, we have 8 objects plotted in Fig.~\ref{fig:sublum} that illustrate this behaviour. 

There will be some over- and under-luminous objects which we cannot distinguish clearly from the MS population and indeed objects falsely labelled as probably under- or over-luminous objects. For example, in the middle panel of Fig.~\ref{fig:sublum}, the distance moduli of the coolest candidate sub-dwarfs seemingly entirely overlaps the upper end of the main stellar locus.  In total 299 objects are tagged as likely to be under-luminous while 344 are tagged over-luminous. This equates to around 5\% of the whole sample going into each category -- a proportion that is a direct consequence of the distance modulus percentile cuts chosen ($95^{th}$ for over-luminous and $5^{th}$ for the sub-luminous). This cut was chosen by visual inspection of the CMD, noting where the sequences show signs of separating.

\cite{MeylanandMaeder1983} estimated a surface density of 10 - 20 blue supergiants (BSGs) per kpc$^{2}$ in the Galactic Plane. Taking our distance range as 3 - 10 kpc we are sampling a projected disk surface area of a little over 9 $\rm kpc^2$. This suggests that our selection of 344 over-luminous stars is too generous by a factor of two or more. Turning to the faint end of the range, we note that \cite{hanetal2003} proposed a space density of sdB stars of $2-4 \times 10^{-6}$ pc$^{-3}$. Assuming that we are detecting sdB stars at distances between 1 to 2.5 kpc then we are sampling a volume of around $\sim0.06$ kpc$^3$ which predicts around $\sim$180 sdB stars. Again this suggests a somewhat over-enthusiastic selection of 344 possible sub-luminous stars. The fact that these selections `over-shoot', suggests that the remaining 90\% is likely to be a relatively clean selection of near main-sequence OB stars.

The candidate stars selected as candidate over- and under-luminous stars are marked as such in the supplementary table (see Table \ref{tbl:photom_params} in the appendix).

\subsubsection{Emission line stars}
\label{sec:emissioncarina}

\begin{figure}
\centering
\includegraphics[width=\columnwidth]{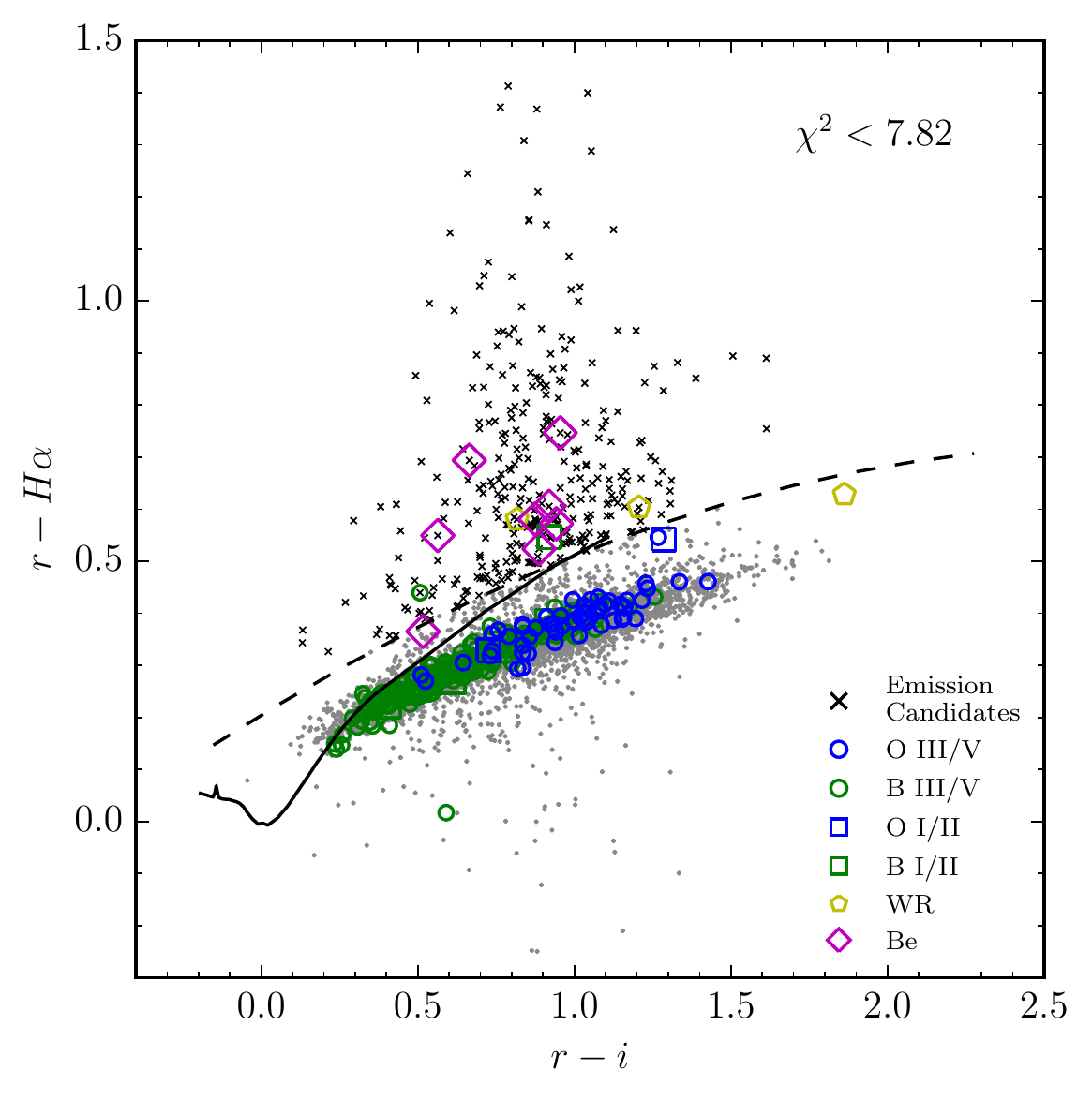}
\includegraphics[width=\columnwidth]{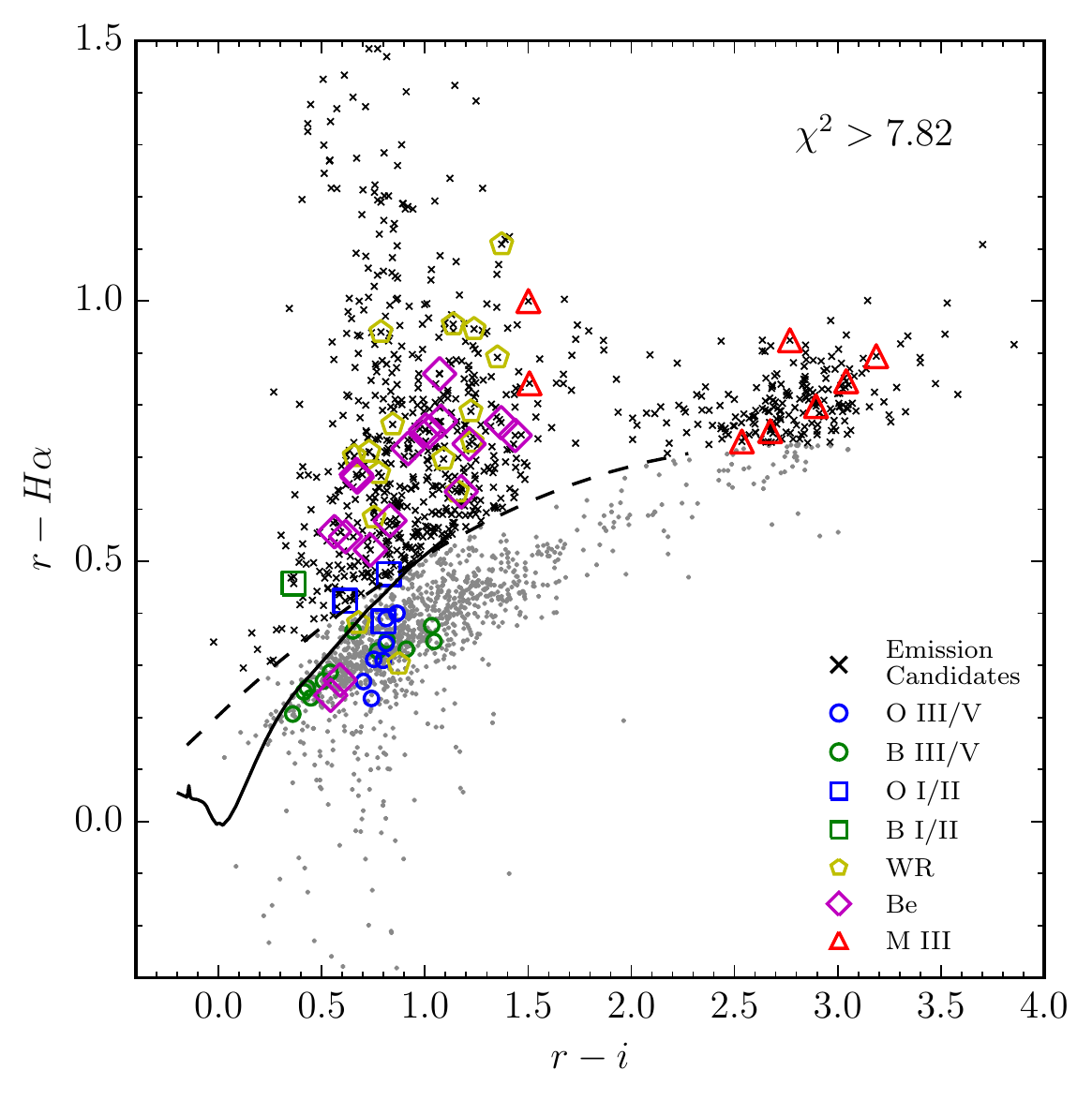}
\caption{Emission line stars selected using the VPHAS+ $(r-i,r- \rm H\alpha)$ diagram with various stars of known spectral class. The top panel shows OB candidates with $\chi^2 < 7.82$ and the bottom panel shows OB candidates with $\chi^2 > 7.82$. Note the wider $r-i$ scale in the bottom panel.}
\label{fig:emission_car}
\end{figure}

We now use the VPHAS+ H$\alpha$ measurements to pick out any emission line stars in our basic blue selection. Following \cite{mohr-smith2015} we use the $(r-i,r- \rm H\alpha)$ diagram to select all objects that lie more than $0.1$\,mag in $r- \rm H\alpha$ above the O9V reddening vector (equating to $\sim 10\AA$ in emission line equivalent width). This secondary selection misses out 152 stars ($\sim1\%$) that are presently without recorded H$\alpha$ magnitudes for quality reasons. Figure \ref{fig:emission_car} shows this selection for objects with $\ltf \geq 4.30$ in the accepted fit group (top panel) and the poor fit group ($\chi^2 > 7.82$, bottom panel). The black crosses show objects selected as emission line stars, while the various coloured symbols show stars with known spectral type from the SIMBAD cross match and those with AAOmega spectra from Section \ref{sec:spectra}. Here we can see that the emission line objects co-locate in the diagram with classical Be (CBe) stars and WR stars while the non-emission line objects co-locate in the diagram with the known dwarf and giant OB stars. Previous experience with the same selection method in the Northern Hemisphere has shown that most candidate emission line stars are CBe stars, at least at brighter magnitudes \citep{Raddi2015}. As discussed in Section \ref{sec:broadclass} the majority of WR and CBe stars produce poor $\chi^2$ values when fit to MS OB star SEDs.

The bottom panel of Fig.~\ref{fig:emission_car} shows in addition a group of poor $\chi^2$ objects with $r-i>2$ that co-locate with a group of known M giant stars. These very red contaminant objects showing extreme $r-i$ values, are found in the bottom right corner of the $(u-g, g-r)$ CC diagram, permitting them to enter into the original selection as candidate highly reddened OB stars. It is likely that the measured $u$ band magnitudes for these intrinsically red objects is affected by red leak discussed by \cite{drewetal2014}, pushing them apparently blue-ward into the OB-star area of the diagram. The fact that none of these objects yields a fit with $\chi^2 < 7.82$ is further illustration of how the SED-fitting is critical in enhancing the overall blue-selection quality.

\begin{figure}
\centering
\includegraphics[width=\columnwidth]{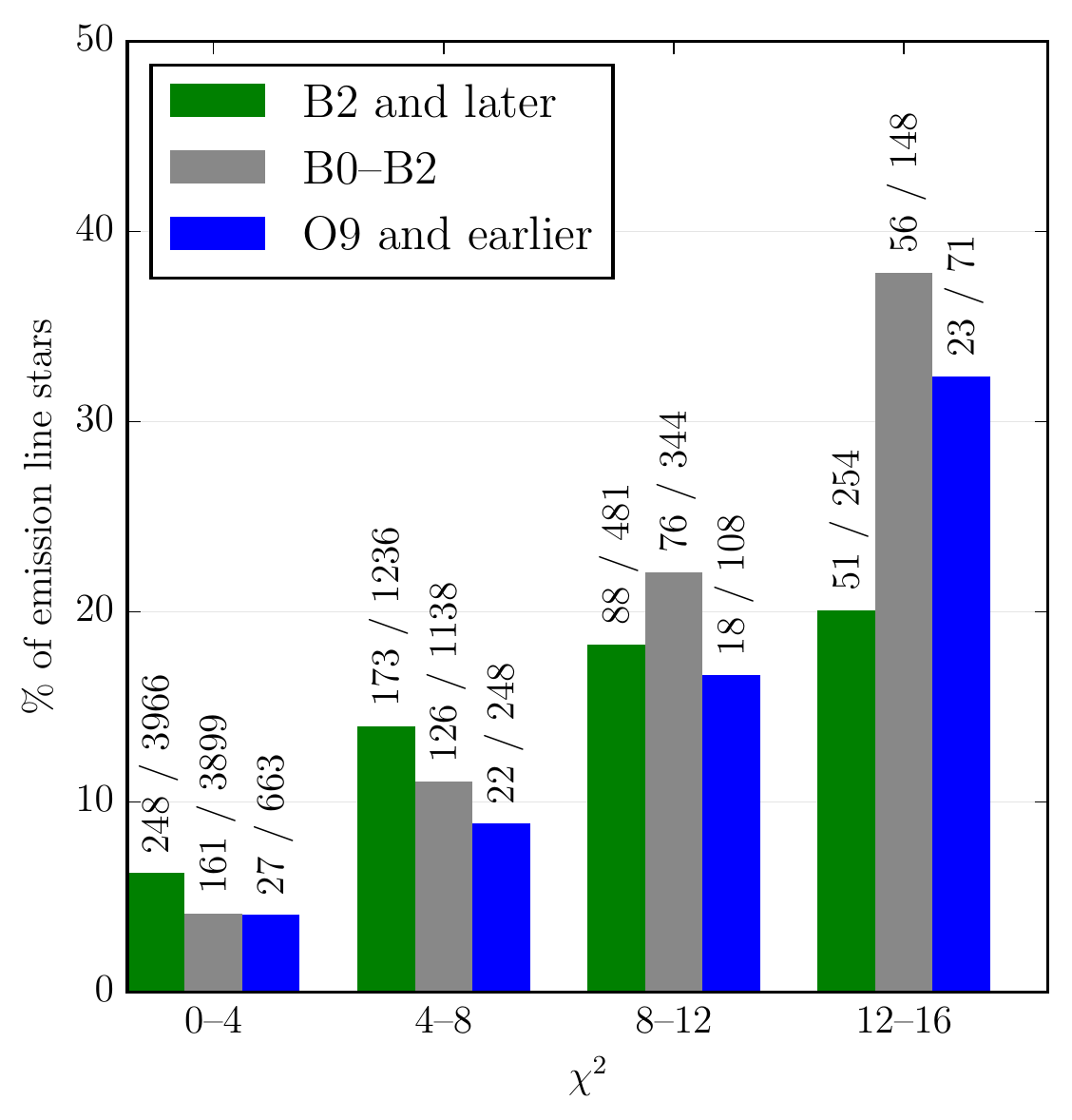}
\caption{The fraction of emission line stars relative to the total number of objects selected in 3 different spectral type groups, as a function of SED fit quality ($\chi^2$). The numbers of objects contributing to each ratio are provided above the coloured bar representing the fraction.  As noted in text, the presence of circumstellar emission in an OB stellar SED unavoidably reduces the quality of fit to a pure photospheric SED.  Hence the fraction of emission line stars rises as $\chi^2$ rises, in all three groups, although the absolute numbers of objects fall. CBe stars and some B[e] and LBV objects most likely dominate the B spectral type range, while WR stars will be most frequent in the O range.}
\label{fig:emission_frac}
\end{figure}


How the candidate emission line stars are distributed in terms of $\chi^2$ and broad spectral type is illustrated in Fig.\ref{fig:emission_frac}.  The growth in the fraction of emission line candidates with declining fit quality stands out.  The numbers provided in the diagram also show that it is the later B group that presents the highest fractions as well as the highest absolute numbers.  Taking all the B stars together with good fits from the first two $\chi^2$ ranges in Fig.~\ref{fig:emission_frac}, the computed fraction of emission line stars is 7.0\%.  This rises to 8.0\%, if $\chi^2$ up to 12 is accepted also.  Given that LBV and B[e] stars are rolled in, these overall fractions certainly appear consistent with the findings of \cite{mcswaingies2005}, as summarised in Fig.5 of their paper.  Our estimates cannot support the 10--20\% range that emerged from the study by \cite{zorecbriot1997}.  The differences here may well reflect the way in which CBe stars are distinguished from normal B stars.  We place a clear threshold in requiring evidence of an H$\alpha$ colour excess equivalent to at least 10~\AA\ emission equivalent width across an observed population -- an approach that is similar to that of \cite{mcswaingies2005} who studied southern clusters.  In older works, quantitative thresholds of this kind were less commonly applied.

Objects with H$\alpha$ excess are marked as such in the database provided in the supplementary materials. Objects with H$\alpha$ emission are removed from the later discussion in section 5 regarding extinction as the derived extinction parameters may be affected by the presence of a circumstellar disk.

\section{Direct spectroscopic validation of the photometric selection and
  SED fits}

\label{sec:spectra}

We now present the new spectroscopy, already used to supplement the known SIMBAD-listed objects in Table~\ref{tbl:carina_breakdown2}, in order to appraise the quality of outcome from our photometric selection and SED fitting procedures.

\subsection{Spectroscopic observations}

Low resolution spectra of 323 OB candidates ($13 < g < 19$) from our selection were taken using the multi-fibre spectrograph, AAOmega, at the Anglo-Australian Telescope, in service mode during June and July 2014. The targets observed were drawn at random from across the $\chi^2$ and effective temperature range (i.e. some poor fits were included).  The large white circle in Fig.~\ref{fig:carinaregion} shows the AAOmega two-degree field of view centred on RA 10 28 45.25 DEC -58 25 56.12 (J2000) containing all observed targets.  

The 580V grating was used, covering the wavelength range $\sim3000 - 6000 \rm\AA$ at a resolution of R$=$1300 in order to capture the critical blue hydrogen and helium photospheric lines needed for reliable classification. The `red' arm grating, chosen mainly for a different programme, was the 1700D: this captured the calcium triplet region which provided some further classification evidence. We used dedicated sky fibres to support pipeline sky subtraction. The data were extracted and reduced to one-dimensional form using the 2dfdr software package with default settings; the wavelength calibration was obtained using a third order polynomial fit. Due to the large magnitude range of the candidates the observations were split into two configurations; a faint set-up and a bright set-up. The faint set-up included objects in the magnitude range $16 \leq g < 19$ and was exposed for a total of 140 minutes. The bright set-up included objects in the magnitude range $13 \leq g < 16$ and needed only 40 minutes exposure. Some of the spectra had to be discarded due to inadequate counts for classification purposes. The 276 spectra retained span a range in signal-to-noise ratio from 20 to 50 with a median value of 35 as measured in the wavelength range $4560 - 4650\AA$.

Each spectrum was continuum fitted and normalized using a spline fit in the \textsc{pyraf} package \textsc{onedspec}. Before further analysis, the prominent diffuse interstellar bands (DIBs) at $\lambda \simeq 4430$, 4892, 4748 and 5362$\AA$ were cut out.

\subsection{Broad classification of the spectra}
\label{sec:broadclass}
The `blue' and `red' wavelength range of each stellar spectrum was visually inspected in order to place the objects into broad spectral types.
Of the 276 candidates observed, using the spectral features listed below, we found that the spectral classes break down as follows:

\begin{itemize}
\item[--]{196 B stars} (HeI absorption lines)
\item[--]{29 O stars} (HeI \& HeII absorption)
\item[--]{27 CBe stars} (H$\beta$ emission \& HeI absorption \& double peaked Paschen)
\item[--]{22 A/F/G stars} (G-band, calcium triplet absorption)
\item[--]{2 WR stars (previously known)}
\end{itemize}

Including the more exotic CBe and WR stars among the desired massive star selection, we find the remaining contamination comes from just a handful of lower mass A, F and G type stars. Subsequent close inspection of the original VPHAS+ data revealed that all of the A/F/G stars had photometry that was compromised either by source de-blending failures, or by poor background subtraction due to a very bright neighbouring star, or by falling close to the edge of a CCD (corrupting the photometry in one or more bands).  

\begin{figure}
\centering
\includegraphics[width=\columnwidth]{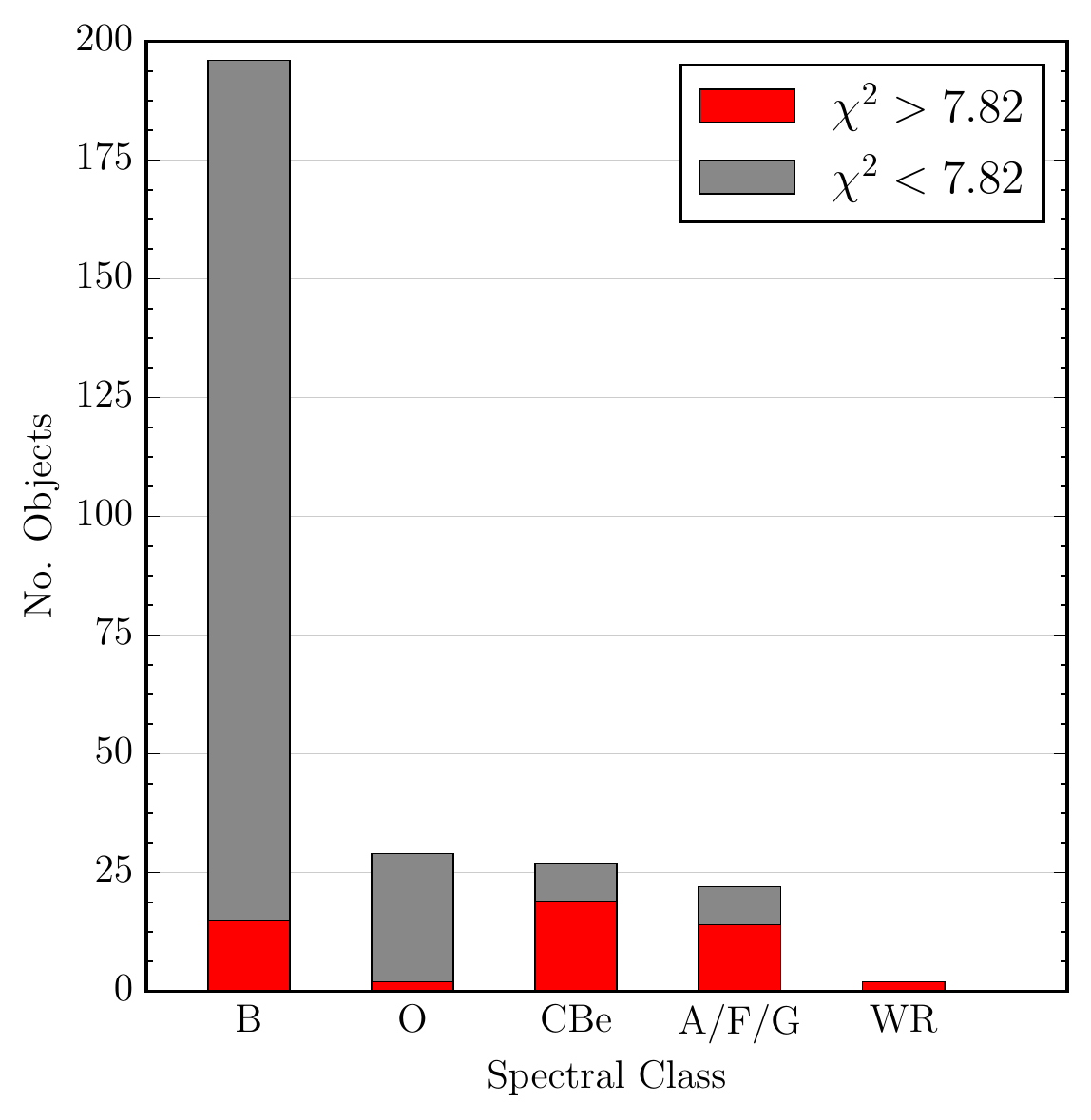}
\caption{The distribution of broad spectral classes found for the sample of stars with AAOmega spectra.}
\label{fig:spectypes}
\end{figure}

Fig. \ref{fig:spectypes} shows the distribution of spectral classes for the selection. The distribution in red is for those objects that have unacceptable $\chi^2 > 7.82$ photometric SED fits. Here we can see that the majority of A/F/G star contaminants have emerged with poor fits to main-sequence OB star SEDs thanks to the contaminated photometry. For studies focused on small numbers of objects, it is quite easy to limit contamination of the selection by visually inspecting thumbnails of every star in the sample. However, when the numbers grow this becomes increasingly less practical and the benefit becomes less obvious given the small percentage of contaminants likely. The SED fitting procedure usefully reduces the contamination rate from $8\%$ to just $3\%$ when making a cut at $\chi^2<7.82$.

We also note that around $70\%$ of the CBe stars are found to have poor $\chi^2$ values. Although the OnIR SEDs of classical Be stars are not greatly different from normal B stars of similar effective temperature, the presence of a warm circumstellar disk can affect the NIR colours. It is in fact easy to separate CBe stars from the rest of the population through detection of line emission via the VPHAS+ narrow band H$\alpha$ filter (see Section \ref{sec:emissioncarina}). Both of the previously known WR stars have similar colours to MS OB stars in the optical and are hence in our selection, but they do not fit well to MS OB star OnIR SEDs, again because of NIR excess -- this time due to their dense stellar winds \citep{Fahertyetal2014}. Inspection of the SED fit residuals for spectroscopically-confirmed OB stars with $\chi^2 > 7.82$ reveals that most of them have been affected by blending in the NIR photometry. At these longer wavelengths, blending will be more common due to the combination of higher prevailing stellar densities and the lower angular resolution of 2MASS.

\subsection{Model atmosphere fitting}

Each OB-star spectrum covering the wavelength range $\sim3000 - 6000 \rm\AA$ (225 stars in total) was fitted to a grid of model spectra in order to derive effective temperatures for comparison to our photometric estimates. The method used for the O stars had to differ from that applied to the B stars in order to adapt to different model grids, supplied in different forms. For the B stars the TLUSTY NLTE grid \citep{LanzandHubeny2007} was used and for the O stars, the more densely sampled FASTWIND grid tailored for these hotter objects \citep[based on,][]{simondiaz2011, pulsetal2005}. As the full spectrum was supplied in the TLUSTY grid, the entire spectrum was used in the fitting procedure. The FASTWIND grid treats individual line profiles, including the hydrogen and helium lines needed to determine effective temperature and surface gravity -- just these (stronger) lines were input to the fitting procedure. Solar metallicity was adopted in both cases. Since this sight line samples only a limited range of Galactocentric radii ($R_G \sim7-10$ kpc) this should be a reasonable approximation.

\subsubsection{B star model fitting}

We begin with a set of TLUSTY model spectra which are parametrised in terms of just effective temperate and surface gravity. Each spectrum in the grid was rotationally broadened to a range of output projected rotation velocities ($v\sin i$) and convolved to match the resolution of the instrument using the \textsc{python} package \textsc{PyAstronomy}. The grid was then linearly interpolated to form a continuous set of models for MCMC sampling, growing the parameter set to three:

\begin{equation}
\theta = \lbrace T_{\rm eff},~ \log(g), ~ vsini \rbrace
\end{equation}
The best fit parameters were derived using MCMC sampling with a $\chi^2$ likelihood function akin to that used in the photometric fits:

\begin{equation}
\label{eqn:chispec}
P \left( SPEC_{obs} \mid \theta \right) \propto \exp\left(-\frac{1}{2}\sum\limits_{i}^{n} \frac{(f(obs)_i - f(mod)_i)^2}{\sigma_i^2}\right)
\end{equation}
where $P \left( SPEC_{obs} \mid \theta \right)$ is the probability of obtaining the observed spectrum ($SPEC_{obs}$) given a set of parameters $\theta$ and $f(obs)_i$ and $f(mod)_i$ as the observed and model normalized flux at each wavelength $i$.

We compare the observed spectra to a grid of parametrised TLUSTY models using uniform priors:

\begin{multline}
\label{eqn:priorsspecB}
P(\theta) = \begin{cases} 1 \quad \text{if} \,\,\begin{cases} 15kK \,\, \leq \,\, T_{\rm eff} \,\, \leq \,\, 30kK \\
1.75 \,\, \leq \,\, \log(g) \,\, \leq \,\, 4.75 \\
0 kms^{-1} \,\, \leq \,\, vsini \,\, \leq 600 kms^{-1} \\
 \end{cases}\\
 0 \quad \text{else}
 \end{cases}
\end{multline}

The boundaries of the priors match the limits of the model grid.

\subsubsection{O star model fitting}

\begin{figure*}
\centering
\includegraphics[width=\textwidth]{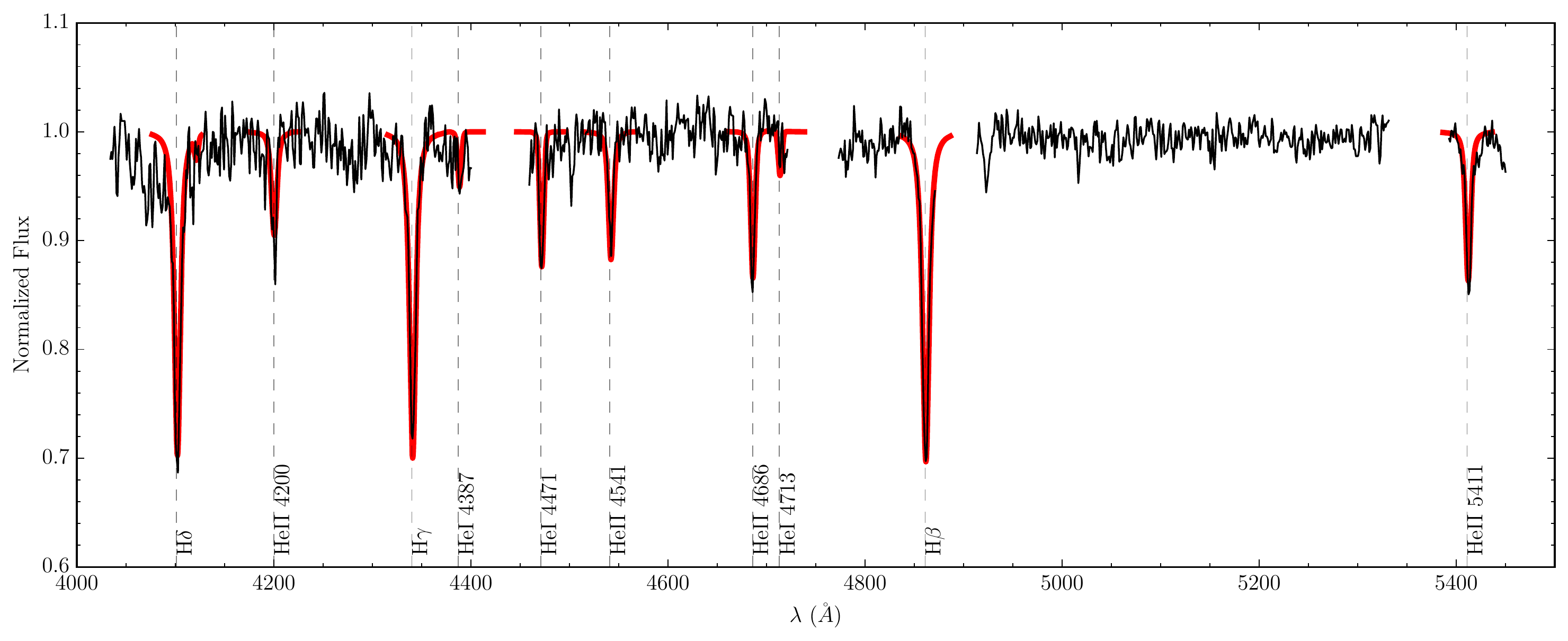}
\includegraphics[width=\textwidth]{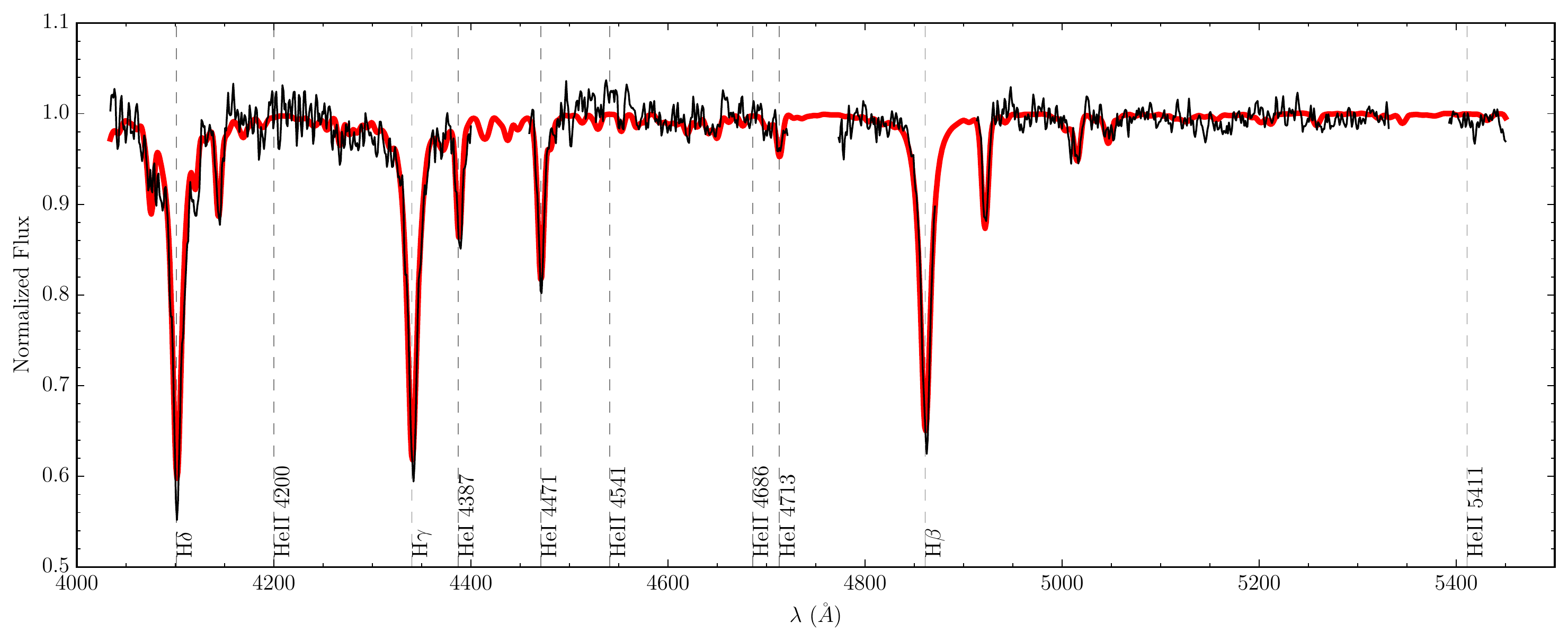}
\caption{Example of spectra and best fit models. Top: ID \#1374, $g=15.5$ mag, $A_0=5.6$, $T_{\rm eff}=38600$ K and $log(g) = 4.19$, with the FASTWIND fit to the hydrogen and helium lines superimposed in red. Bottom: ID \#2593 $g=17.2$ mag, $A_0=4.5$, $T_{\rm eff}=23100$ K and $log(g) = 4.32$, with the TLUSTY NLTE full spectrum fit superimposed. The signal to noise ratios of the two spectra are $\sim25$ (top) and $\sim31$ (bottom).
}
\label{fig:spectra}
\end{figure*}

We deploy a set of FASTWIND models that consist of the following line profiles: H$\delta$, H$\gamma$, H$\beta$, HeI 4387, 4471, 4713, HeII 4200, 4541, 4686, and 5411.

The models are parametrised by the following:

\begin{itemize}
\item[--]{$T_{\rm eff}$: effective temperature}
\item[--]{$\log(g)$: surface gravity}
\item[--]{$n(He)$: helium abundance, by number}
\item[--]{$\xi_t$: micro turbulence}
\item[--]{$\rm Q$: wind strength parameter, $\dot{M}(R~v_{\infty})^{-1.5}$}
\item[--]{$\beta$: exponent of the wind velocity law}
\item[--]{$v\sin i$: rotational velocity}
\item[--]{$\rm RV$: heliocentric radial velocity}
\end{itemize}

In this case each line profile was fit independently using the same style of likelihood function as Eqn. \ref{eqn:chispec}. The overall function is the product of probabilities for all line profiles:

\begin{equation}
\label{eqn:prod}
P ( SPEC_{obs} \mid \theta) \propto \prod\limits_{j}^{n} P ( LP_{obs(j)} \mid \theta)
\end{equation}
where $P ( SPEC_{obs} \mid \theta)$ is the probability of obtaining the observed spectrum ($SPEC_{obs}$), given a set of parameters $\theta$, and where $P ( LP_{obs(j)} \mid \theta)$ is the probability of obtaining the observed line profile, $LP_{obs(j)}$, again given a set of parameters $\theta$.

The parameters $\xi_t$ and $\beta$ were fixed at typical values, 5 $kms^{-1}$ and 0.8 respectively, as our spectra are too low resolution to detect any tangible change in them. As for the B stars, each spectrum in the grid was rotationally broadened to deliver a set of $v\sin i$ templates, which were all subsequently convolved with a Gaussian to match the resolution of the instrument.  Again, because of the limited resolution of the spectra, the contribution to line profiles from macroturbulent broadening, known to affect O stars \citep{simondiazandherrero2014, markova14}, can be regarded as subsumed within the typically more prominent rotational broadening.  This leaves 6 free parameters:

\begin{equation}
\theta = \lbrace T_{\rm eff},~ \log(g), ~ vsini, ~ RV, ~ \log(Q), ~ n(He) \rbrace
\end{equation}

The uniform grid of templates was linearly interpolated to form a continuous grid for MCMC sampling. The following uniform priors for each parameter were adopted and are defined by the limits of the model grid:

\begin{multline}
\label{eqn:priorsspecO}
P(\theta) = \begin{cases} 1 \quad \text{if} \,\,\begin{cases} 25kK \,\, \leq \,\, T_{\rm eff} \,\, \leq \,\, 55kK \\
3.0 \,\, \leq \,\, \log(g) \,\, \leq \,\, 4.4 \\
0 kms^{-1} \,\, \leq \,\, vsini \,\, \leq 600 kms^{-1} \\
-400 kms^{-1} \,\, \leq \,\, RV \,\, \leq 400 kms^{-1}\\
-14  \,\, \leq \,\, \log(Q) \,\, \leq -12.5\\
\,\,0.06 \leq \,\, n(He) \,\, \leq 0.20
\\
 \end{cases}\\
 0 \quad \text{else}
 \end{cases}
\end{multline}

The key constraint expected from these fits is on effective temperature. There is also useful information contained in the derived surface gravities and rotational velocities. Given the low resolution nature of the spectra we treat all the other parameters as `nuisance parameters' since the derived values will carry large uncertainties and are not relevant to this discussion. However, their influence on the other parameters is, by default, taken into account when using the marginalised posterior probability distributions. We also tried fitting the O stars to O star NLTE TLUSTY models \citep{LanzandHubeny2003} using the same method as for the B stars and found little difference in the average derived temperatures and surface gravities compared with the FASTWIND results: we note there is a tendency for the TLUSTY models to offer combinations of slightly higher surface gravities and effective temperatures relative to FASTWIND, especially in the hottest cases (see the example discussed below in the next section).

\subsection{Results of the spectroscopic fitting}

The median of the posterior probability distributions for each parameter is taken as the best fit value. The $16^{th}$ and $84^{th}$ percentiles are taken as the upper and lower uncertainty on each parameter. The best-fit spectroscopic parameters for all the OB stars observed with AAOmega are added to the database provided as supplementary material.  Fig. \ref{fig:spectra} shows example spectra, after normalisation and excision of DIBs present, over-plotted on the best fit model. The top panel presents an O star spectrum compared with FASTWIND best-fit line profiles, while the bottom panel presents a comparison between an observed B-star spectrum with an NLTE TLUSTY fit.

\begin{figure}
\centering
\includegraphics[width=\columnwidth]{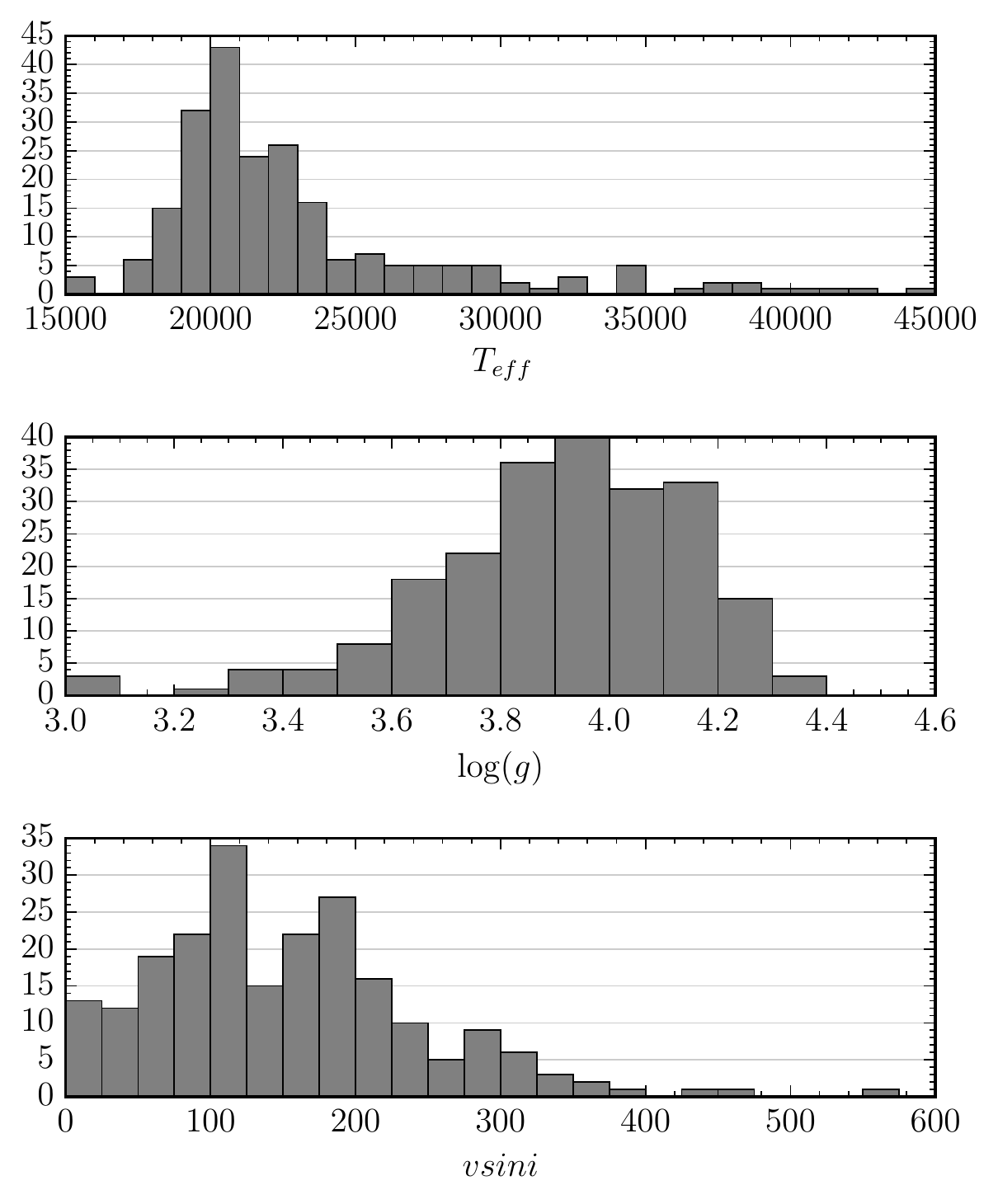}
\caption{The distributions of effective temperature, surface gravity and projected rotational velocity obtained from the spectroscopic fits.}
\label{fig:spec_hist}
\end{figure}

Fig. \ref{fig:spec_hist} shows the distribution in $T_{\rm eff}$, $\log(g)$ and $v\sin i$ for the spectroscopic fits of OB stars. The distribution in $T_{\rm eff}$ peaks at around 20kK and falls off at higher temperatures. This is very similar to the distribution in the photometric temperature estimates shown in \cite{mohr-smith2015}. As expected the majority of stars show MS surface gravities with a median value a little less than 4 (more precisely, $3.92$). The three lowest surface gravity objects $\log(g) \lesssim3.2$ are likely to be evolved B stars. Their relatively cool temperatures ($19 \rm kK < T_{\rm eff} < 24 \rm kK$) and lower surface gravities would be consistent with luminosity class II bright giants \citep{Schmidt-Kaler1982}. Their over-luminous position in the $(g,u-g)$ CMD also supports the notion that these are evolved objects (see Section \ref{sec:sublum}: these are 3 of the objects emphasised in black in Fig.~\ref{fig:sublum}).

One very hot O star whose spectrum has been compared with model atmospheres, has also been selected as a subdwarf (ID 1558) using the technique outlined in Section \ref{sec:sublum}. We found that the fits both to the FASTWIND grid and to the TLUSTY (OSTAR) grid struggled with the upper limits on $\log g$ (respectively 4.3 and 4.75: the fits returned gave $\log g = 4.2$ and 4.7).  To overcome this problem, we carried out a fit against a grid of hydrogen and helium NLTE model atmospheres used by Napiwotzki (1999) for the analysis of central stars of planetary nebulae (CSPN). The fit results are $T_{eff} = 55\pm2$~kK, $\log g = 4.4\pm 0.2$ and $n(\mathrm{He}) = 0.23\pm0.05$.  Whilst the neglect of metals in this third model grid might cause a modest systematic shift, a gravity well below what is typical for most hot sdO stars \citep[$5.5 \lesssim \log g \lesssim 6.0$][]{greenetal2008} is supported. This, and the presence of weak, spatially-unresolved [OIII]~$\lambda\lambda$4959,5007 in emission in this object's spectrum suggests post-AGB status for this star.  Interestingly, in addition, the sdO atmosphere exhibits helium enrichment $n(\mathrm{He}) = 0.23$ and 0.15, respectively, from the CSPN and FASTWIND grids).  The history of this object could be similar to that of the unusual PN central star, K648 in M15 \citep{rauchetal2002}, which shows signs of material being mixed into the atmosphere during a final helium flash.  For now we acknowledge that our one `sub-luminous' spectroscopic target is potentially not so sub-luminous and that higher quality data are required to conclude on its nature.  Nevertheless, the placement of this star in the non-massive ``subluminous'' group is sound.

The distribution in projected rotational velocities ($v\sin i$) for the entire sample of OB stars is broadly consistent with the distributions presented by \citet{Huangetal2010}, for B stars, and by \cite{simondiazandherrero2014}, describing O stars in the Milky Way.  The values returned for this parameter are provided in the database for individual objects for completeness. But we caution that as individual measures they will be appreciably more uncertain than the quoted MCMC-derived random errors might suggest.  This is due to limited spectral resolution (corresponding to a velocity resolution of $\sim$230 km~s$^{-1}$, and the moderate S/N of the data (most often, $30-40$)-- which have both a direct impact through e.g. uncertainties in continuum placement, and more indirectly through the lack of capacity to distinguish binarity (SB2 component blending).

\subsection{Comparison of the spectroscopic and photometric best-fit effective temperatures} 

\begin{figure}
\centering
\includegraphics[width=\columnwidth]{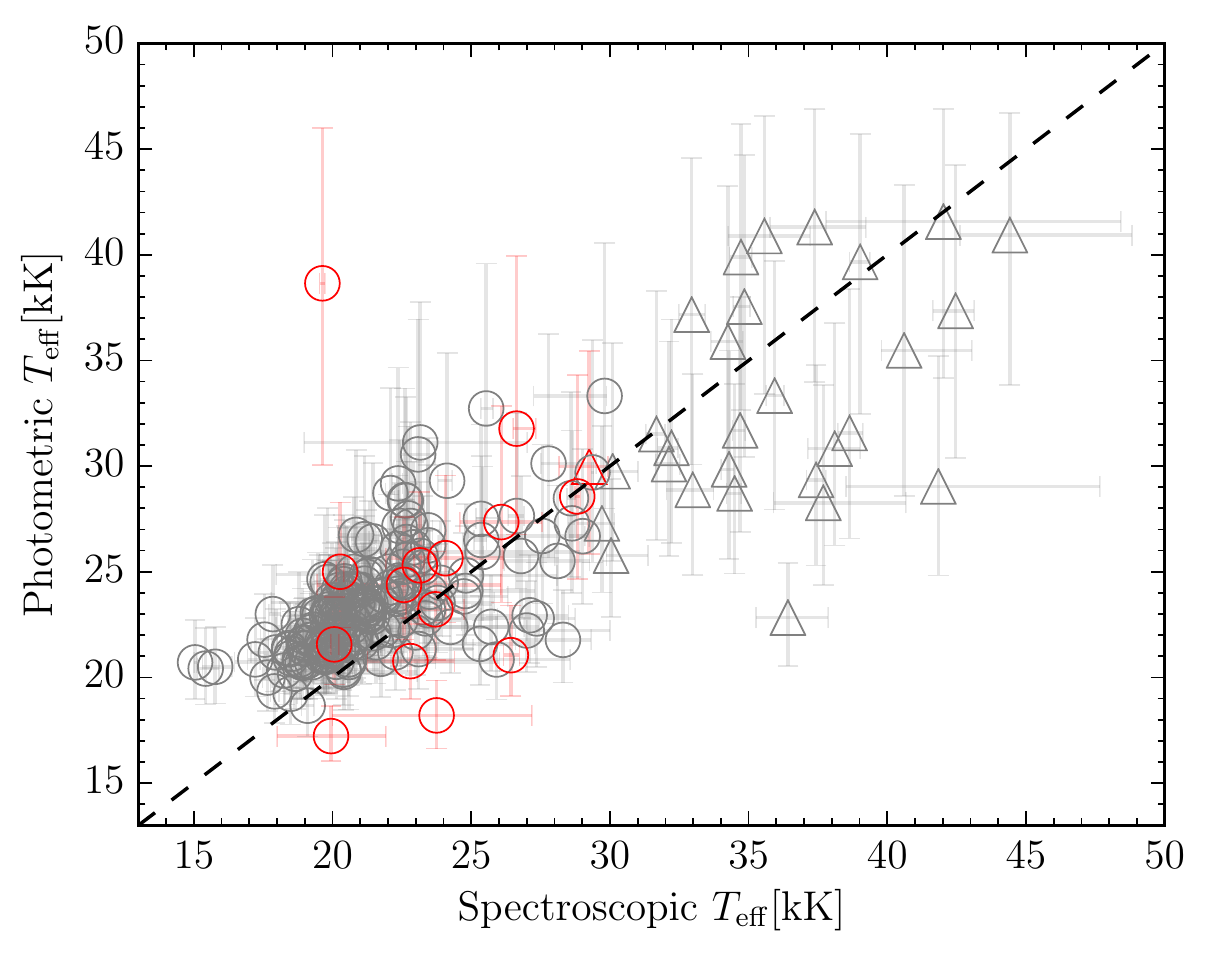}
\caption[Comparison of the derived effective temperature from the photometric SED fitting with those derived from spectroscopy.]{Comparison of the derived effective temperature from the photometric SED fitting with those derived from spectroscopy. The triangles are O stars and the circles are B stars. Symbols coloured in red had photometric SED fits with $\chi^2>7.82$.}
\label{fig:specteffcompare}
\end{figure}

\begin{figure*}
\centering
\includegraphics[width=\textwidth]{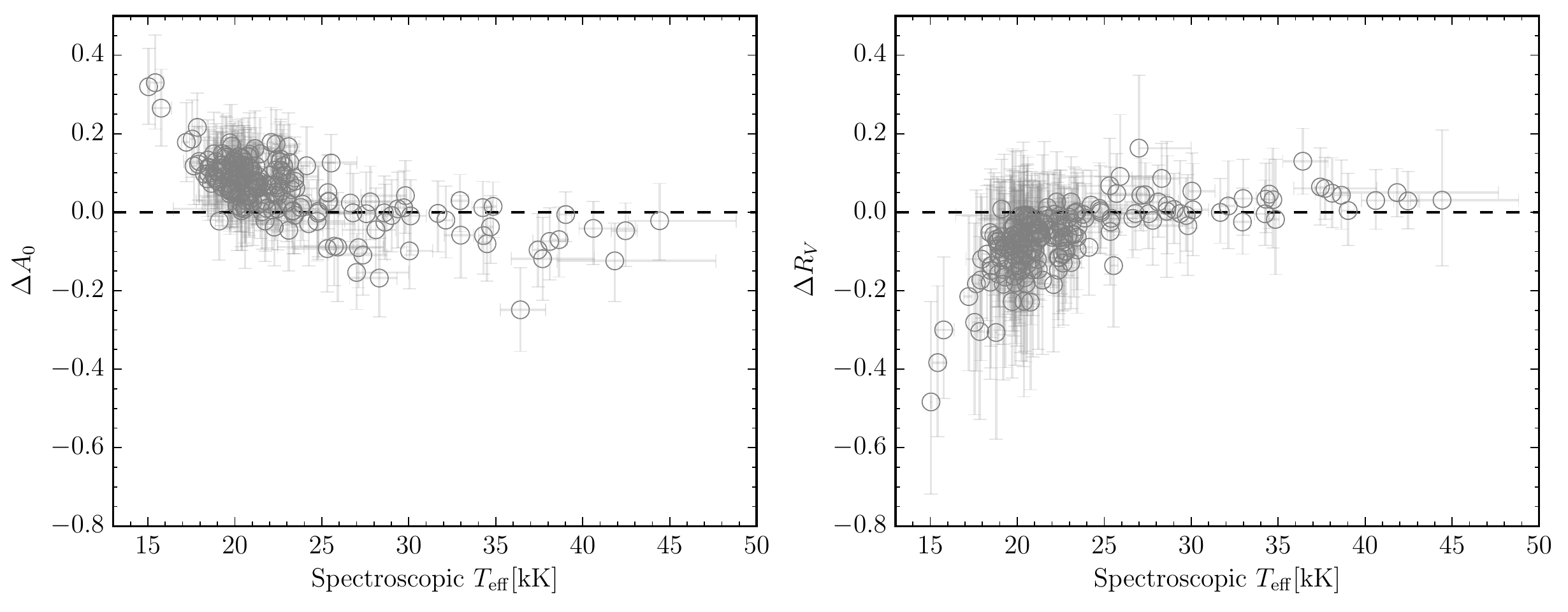}
\caption{Change in derived reddening parameters when using a temperature prior defined by the spectroscopic fits (free - fixed).}
\label{fig:fixedteffcompare}
\end{figure*}

Figure \ref{fig:specteffcompare} shows the effective temperatures determined from the photometric SED fits compared to those determined by spectroscopy. The red circles show objects with $\chi^2 > 7.82$ in the SED fitting routine. We find that, despite the necessary reservations about determining effective temperatures through photometry alone, there is a good correlation between the two methods. We do however see a systematic tendency in the photometric method to under-estimate the temperature of O stars and over-estimate those of the B stars, when compared to the spectroscopy. As the photometric calibration appears to be well behaved, this is most likely to be due to differences between the photometric effective temperature scales in the Padova isochrones (based on LTE model atmospheres, Bressan et al 2012) and those relevant to the TLUSTY and FASTWIND NLTE models used to fit the spectra. Another factor that affects the extreme cases is photometric blending or contamination. For example the poor $\chi^2$ object with the largest disagreement is close to a very red object which is very likely to have affected the measured magnitudes, particularly 2MASS $J$ to $K$.

With the extra information on temperature provided by the spectroscopic fits, we can check what effect the systematic offset in the photometrically derived $\ltf$ has on the derived reddening parameters. To do this we re-calculated the SED fits using a much narrower uniform prior on temperature, based on the 16th and 84th percentiles of the marginalized posterior distribution of spectroscopic temperature for each star (as indicated by the horizontal error bars in Fig. \ref{fig:specteffcompare} \& Fig. \ref{fig:fixedteffcompare}). Fig. \ref{fig:fixedteffcompare} shows the impact on the derived values of $A_0$ and $R_V$ when the more precise `restricted' temperature prior is used in place of almost no constraint in the SED fitting. Here we can see that for both reddening parameters that restricting $\ltf$ has a greater effect on stars below $\sim$25kK. This is due to the higher sensitivity of the shape of the SED as a function of temperature for the cooler stars. At effective temperatures above $\sim25$kK, the SEDs approach the Rayleigh-Jeans limit more closely, lowering the sensitivity to temperature: in this regime, the shape of the SED is almost entirely dominated by extinction.  

It is important to notice that these systematic effects are very small; for stars with $T_{\rm eff} \geq 25$kK the median difference in $A_0$ and $R_V$ (`free' - `restricted') is respectively -0.025 and 0.026, and for those with $T_{\rm eff} < 25$kK the median differences rise to 0.09 and -0.07. In the worst cases these systematic uncertainties are comparable to the random uncertainties on the parameters due to the photometric errors. However, whilst there is indeed a modest disagreement between the photometric and spectroscopic effective temperature scales, it would amount to forcing photometric fits to systematically `incorrect' SED models if the temperature prior were restricted to suit the spectroscopic results. This idea is supported by systematically poorer fits in the `restricted' case where the median $\chi^2=2.99$, in contrast to $\chi^2=1.58$ in the `free' case.  Accordingly it does not make sense to regard the trends seen in Fig.~\ref{fig:fixedteffcompare} as providing a systematic correction to the parameter estimates based on photometry.  What it does all point to is the issue of fit degeneracy between effective temperature and extinction: this begins to acquire some significance towards the cool end of the OB range where we find the larger discrepancies.  But the magnitude of this effect is very much smaller than that affecting fits to broad-band photometry of A to K type stars (see Bailer-Jones, 2011), underlining the long-recognised value of OB stars in tracing Galactic extinction. 

\section{Extinction properties and spatial distribution}

It was demonstrated by \cite{mohr-smith2015} and also in preceding sections that the SED-fitting process places strongest constraints on the reddening parameters, $A_0$ and $R_V$, achieving typical precisions in the region of $\pm0.1$.  These are now put to use to better understand the full pencil beam, passing through the Carina Arm to distances at least as great as $\sim$7~kpc, the distance to NGC 3603.  We begin with an overview of extinction across the region, before extracting more detail. 

In this analysis and discussion, we remove from consideration the objects tagged as either emission line stars or as potentially sub-luminous (reducing 5915 objects to 4770).

\subsection{Overview of extinction in the region}
\label{sec:extlaw}

It is already well known, going back in the literature to before Cardelli et al (1989), that the Carina region is associated with flatter extinction laws, charaacterised by higher $R_V$ $(> 3.5)$.
\cite{fitzpatrickandmassa2007} and \cite{huretal2012} found this near the open clusters Trumpler 16 (Tr 16) and NGC 3293, whilst normal values of $R_V\sim3.1$ still seem to apply to less-reddened stars with $A_0 \lesssim 2$ magnitudes.  In a Chandra X-ray study of the wider Carina Nebula, spanning a sky area of 1.49 sq.deg., Povich et al (2011) confirmed this general pattern and recommended $R_V = 4$ as representative of higher extinctions (from $A_V \simeq 1.8$ to $A_V \sim 4$, based on an X-ray selected sample of 94 OB candidates).

\begin{figure} 
\centering
\includegraphics[width=\columnwidth]{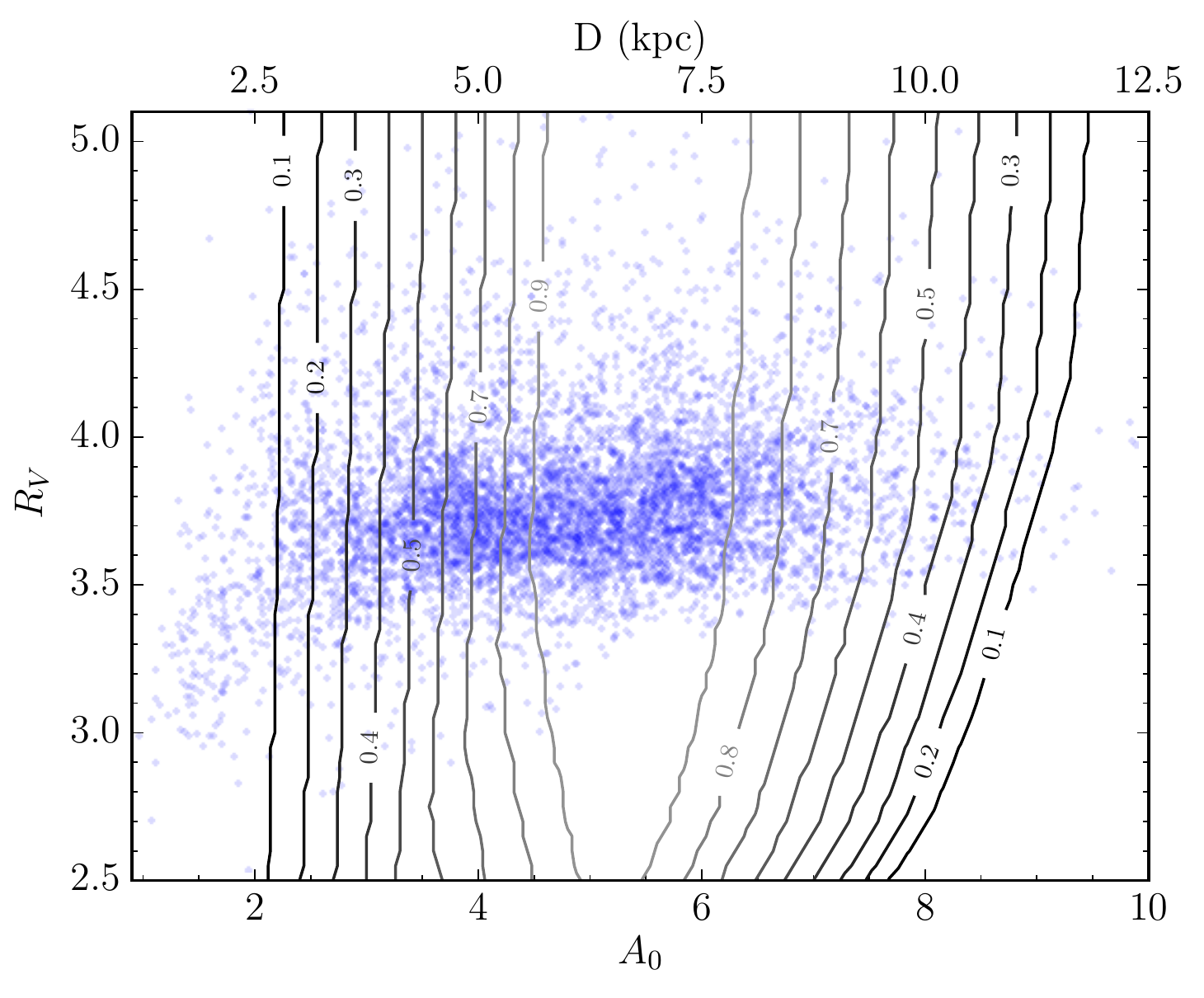}
\caption{Contour lines estimating the OB star detection fraction in VPHAS+ for different combinations of $A_0$ and $R_V$ superposed on the obtained distribution (shown in blue). An increase in extinction of 0.8 mag/kpc is adopted in the modelling of the contours -- the implied heliocentric distance scale is marked across the top. The offset between the range of maximum detectability and the measured extinctions is to be expected without the incorporation of a Galactic model and IMF weighting: the drop off in detections beyond $A_0 \sim 7$ is attributable to the overall downward trend in stellar density.  The pencil beam studied exits the Solar Circle at a distance of $\sim$6~kpc, while at a distance of 10~kpc, the Galactocentric radius is 10--11~kpc.}
\label{fig:a0rvlims}.
\end{figure}

\begin{figure} 
\centering
\includegraphics[width=\columnwidth]{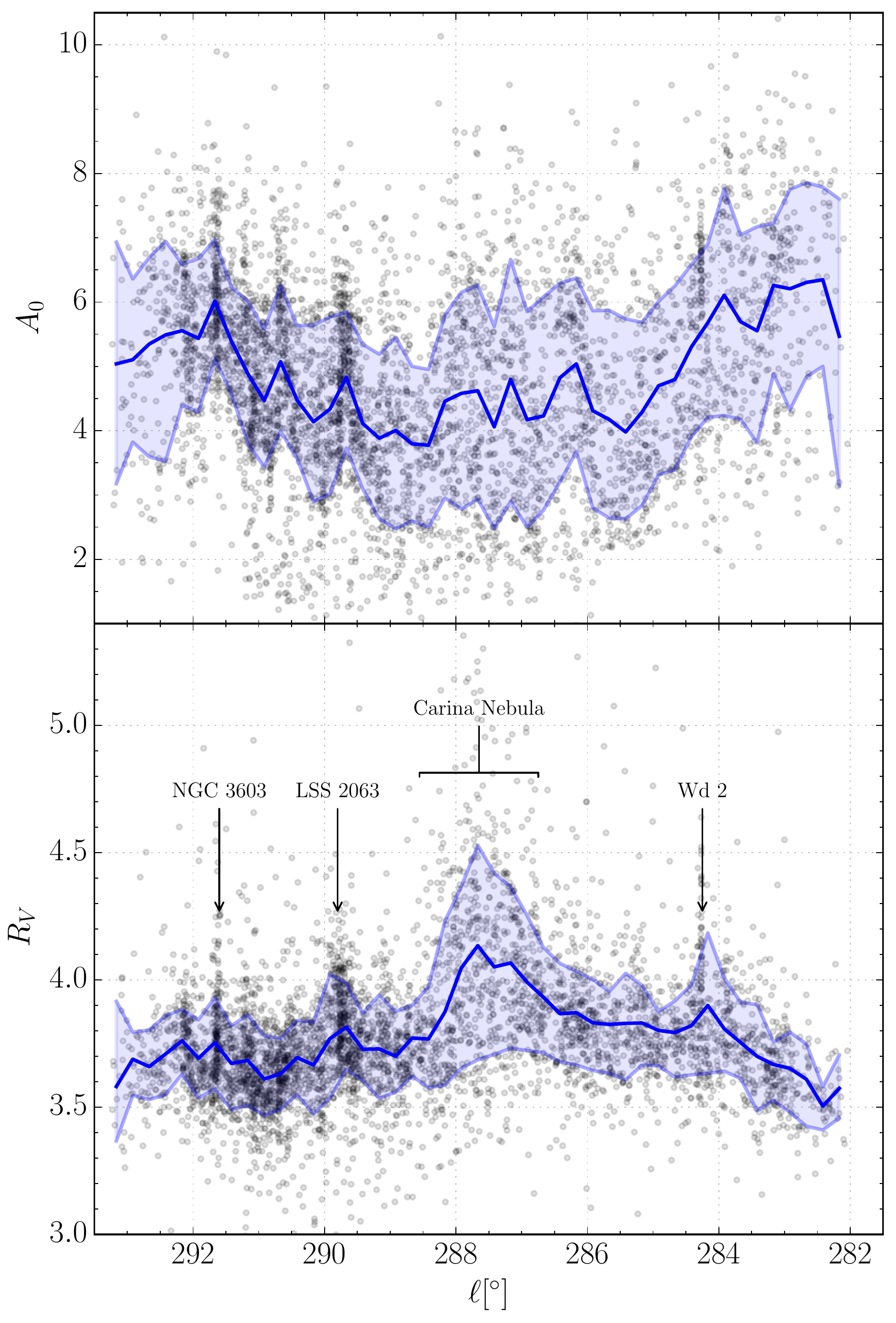}
\caption{The distributions of $A_0$, top panel, and $R_V$ lower panel, as functions of Galactic longitude.  The plot is restricted to non-emission $\chi^2 < 7.82$ candidate OB stars in the Galactic latitude range $0.0^{\circ} > b > -1.5^{\circ}$ (4770 stars).  The blue line, in both panels, is the running mean of the distribution sampled every 0.25 degrees.  The standard deviation about the mean is also shown shaded in light blue.  The typical random error on each data point is <0.1 (magnitudes, in the case of $A_0$).}
\label{fig:a0rv}.
\end{figure}

First, we need to identify the spatial and extinction parameter domain open to examination. To this end, we consider the constraints set by the sample magnitude limits using some straightforward modelling: we take all of the photometric OB main sequence star models in the grid used in SED fitting ($\ltf\geq4.30$) and redden them, assuming a rise in $A_0$ of 0.8~mag/kpc (retrospectively a plausible growth rate for the region, see Section \ref{sec:diffext}) for all values of $R_V$. This permits us to see which combinations of $A_0$ and $R_V$ produce an SED falling within the survey-imposed limits: $12<u<21$, $12<i<21$, $12<r<21$ and $13<g<20$. We collect the fraction of simulated SEDs meeting requirement for each combination of $A_0$ and $R_V$ and (by implication of the adopted $A_0$ rise rate) distance.  This model of detectability does not reveal how many stars we expect to find in the selection given $A_0$ and $R_V$, as this would require taking into account IMF weighting, the growth in the volume captured with increasing distance, and a Galactic stellar density model. Instead, our present purpose is restricted to establishing limits on the detectable ($R_V$, $A_0$) parameter space.

Fig.~\ref{fig:a0rvlims} answers this question: it turns out that the amount of extinction, $A_0$, places the stronger constraints, in that the detectability at $A_0 < 2$ and $A_0 > 10$ of OB stars is low.  Note that only B stars are intrinsically faint enough at the low end of this range to be picked up, while only O stars are luminous enough at the other.  The constraints on $R_V$ within the open $A_0$ window are much weaker: there is little/no constraint on $R_V$ until, at $A_0 > 9$, SEDs presenting $R_V \lesssim 2.5$ become increasingly difficult to capture.  This reflects the link between falling $R_V$ and a steepening extinction law (at higher $A_0$), such that the $u$ magnitude will be more readily pushed beyond the faint limit as $R_V$ drops.

The observed distribution in $R_V$ as a function of $A_0$, superimposed in Fig.~\ref{fig:a0rvlims}, reveals a striking absence of correlation between these parameters, except that only below $A_0 \simeq 2$ is $R_V < 3.5$ relatively common.  From $A_0 \gtrsim 3$ to $A_0 \sim 8$, where the distribution begins to peter out, the trend is flat and broad.  Relatively few objects are picked up at distances below 2.5--3~kpc, indicating this new sample misses out the foreground to the Carina Arm.  The apparent offset between the peak detectability range defined by the contours in the figure ($4 \lesssim A_0 \lesssim 10$), and main concentration of the measurements ($3 \lesssim A_0 \lesssim 7$) has two different origins dependent on Galactic longitude: near the Carina Arm tangent 0.8 mag/kpc is too low a growth rate in extinction, while at greater longitudes where this rate is better matched there is a strong decline in stellar density beyond both the Solar Circle (crossed at a heliocentric distance of 5.5~kpc) and the outer Carina Arm (crossed at 7--10 kpc, depending in detail on longitude).  We now turn to considering the pattern of extinction variation on the plane of the sky to explain why this is so.


Fig.~\ref{fig:a0rv} shows how $A_0$ and $R_V$ vary across the region as a function of Galactic longitude.  The 4770 stars plotted are limited to the latitude range $-1.5^{\circ} < b < 0.0^{\circ}$, the most completely and densely sampled part of the region (see Fig.~\ref{fig:carina_OB}).  In the upper panel it is unsurprising to see the wide spread in $A_0$ at all longitudes, but there is nevertheless the appearance of a broad dip such that the median extinction is 4--5 in the range $285^{\circ} < \ell < 289^{\circ}$, as compared with 5--6 to either side.  This broad minimum is where the relatively nearby Carina Nebula is located.  

At the lower longitudes closest to the tangent region, the $A_0$ distribution is most disorderly and the highest measures are found.  The Planck XI dust map (Abergel et al 2013) indicates that the total dust optical depth at these longitudes is strongly variable and typically $\sim$ twice as high as at $\ell > 285^{\circ}$, implying total visual extinctions of around 15--20 magnitudes.  This is consistent with what is known of the CO distribution \citep{Grabelsky1987}.  In contrast, at $\ell > 289^{\circ}$, the pattern is at its most regular: indeed, there are signs of a gradual rise in extinction with increasing longitude, passing through the location of NGC 3603.  As shall become evident in section~\ref{sec:diffext}, this trend translates into a broadly regular increase in distance sampled with increasing $A_0$.  The OB stars found in this region mainly belong to the receding outer Carina Arm and more nearly sample the total dust column ($\tau_{353} \sim 2\times10^{-4}$ corresponding to a maximum visual extinction of 8--10, see Abergel et al 2013).

\subsection{Variations in $A_0$ and $R_V$ near massive clusters, and a newly-identified OB association}

\begin{figure*}
\centering
\includegraphics[width=0.65\textwidth]{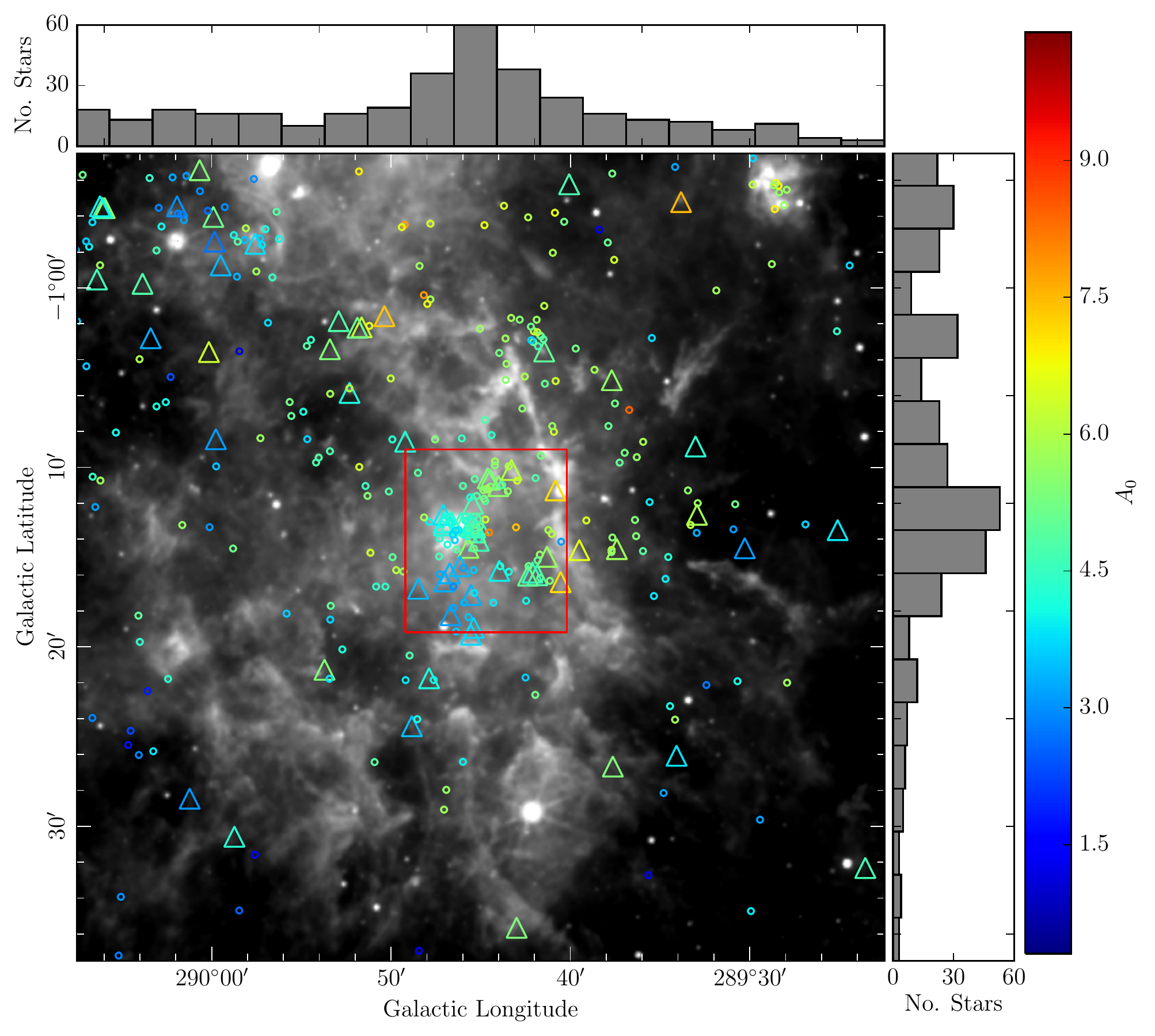}
\caption{A hitherto unrecognised massive OB clustering, scattered around $\ell = 289^{\circ}.77$, $b = -1^{\circ}.22$, the position of the extreme O supergiant, LSS 2063.  High confidence OB candidates are over-plotted on warm dust emission imaged at 12$\mu$m by the WISE mission \citep{Wrightetal2010}.  The histograms above and to the right of the image show the Galactic longitude and latitude distributions of the objects found. Triangle symbols indicate candidate O stars while the smaller circles mark candidate early B stars. Each object is coloured according to best-fit $A_0$. The main concentration of 140 candidates is enclosed in a red box of dimension $0.15\times0.17$ sq.deg.}
\label{fig:newcluster}
\end{figure*}

The most striking feature in the $R_V$ variation with Galactic longitude (lower panel in Fig.~\ref{fig:a0rv}) is the explosion of high values in the mid-longitude range (286.5 to 288, roughly) occupied by the Carina Nebula: any value from $\sim 3.5$ up to more than 5 is apparently possible. Elsewhere excursions beyond $\sim$4 are relatively infrequent.  Interestingly, echoes of this behaviour are also apparent at the longitudes of Westerlund 2 and NGC 3603.  The relatively sharp spikes associated with these two clusters follow from their greater distances and correspondingly reduced angular sizes. Indeed this result suggests that systematically higher $R_V$ and flatter optical extinction laws -- implying larger typical dust grain sizes -- is a persistent property of the environments around young massive (O-star rich) clusters.  This may betray greater columns of dust grain populations biased to larger typical grain sizes in the surrounding molecular gas, unused in the star formation event, or the origin may be more dynamic and arise as a consequence of star formation feedback \citep[see e.g][]{Draine2011,Allenetal2014,Ochsendorfer2015}: further detailed investigation is needed.

In general terms, $3.5 < R_V < 3.8$ appears typical of the Carina Arm field, while near the prominent clusters, average values of $3.7 < R_V < 4.2$ are much more frequently encountered -- with the Carina Nebula presenting as an extreme case.  

On collapsing the Galactic latitude distribution, as in Fig.~\ref{fig:a0rv} (and also in the lower panel of Fig.~\ref{fig:carina_OB}), it becomes easy to pick out the well-known massive clusters -- showing up as dense concentrations, linking to localised peaks in $R_V$ and $A_0$.  In addition, the eye is drawn in both figures to an OB clustering near $\ell \sim 290^{\circ}$ that has no clear literature counterpart, so far.  A close-up of the on-sky distribution of the $\chi^2 < 7.82$ objects in this region, pinpointing the main clustering at $\ell = 289.77^{\circ}$, $b = -1.22^{\circ}$, is provided as Fig.~\ref{fig:newcluster}, with WISE 8$\mu$m (warm dust) emission added as background.  Visual extinctions 4.5--5 magnitudes are typical.  The total number of objects falling in the red box, of rough dimension $9\times10$ sq.arcmin, superimposed on Fig.~\ref{fig:newcluster} is 103, of which 32 are candidate O stars.

The most dense sub-clustering, of 42 objects spread across $\sim$2 arcmin (coloured cyan in Fig.~\ref{fig:newcluster}), coincides with LSS~2063, an 11$^{th}$ magnitude star, classified by \cite{WandF2000} as O5If$+$.  These same authors give $B - V = 1$, which we interpret as a likely visual extinction of about 5 (adopting an intrinsic colour of -0.28 from Martins \& Plez 2006, and $R_V \sim 3.8$, on the basis of Fig.~\ref{fig:a0rv}).  This fits in well with the newly-revealed clustering.  In the first edition of their Galactic O-Star Catalogue, \cite{Maiz-Apellanizetal2004} cautiously described LSS~2063 as a `field' object, there being no clear reports of an associated cluster at the time.  Nevertheless, \cite{dbs2003} did catalogue a near-infrared cluster essentially co-incident with LSS~2063, and more recently \cite{Majaess2013} has noted this location as a plausible YSO clustering based on raised mid-infrared emission (Majaess 133, in the associated catalogue).  \cite{Georgelin2000} noted the HII emission here, in Gum 35 (also known as RWC 54a), and cited the \cite{caswellandhaynes1987} HI recombination line measurement that places the region at the same kinematic distance as NGC 3603 ($\sim$7 kpc away). \cite{Avedisova2002} in her list of star forming regions also notes a coincidence between radio emission in this location and RAFGL 4120 (an earlier mid-IR detection effectively).  So here, finally, the optical has caught up and the very much larger group of ionizing stars, helping LSS~2063 shape this environment, has emerged.

At a distance of about 7~kpc, the over-density's full angular extent of 9-10 arcmin corresponds to a linear size of $\sim$20~pc, allowing this clustering to be viewed as an OB association.  The count of 103 OB stars misses LSS 2063 itself and 4 more that are also above the VPHAS+ bright limit (an O9.5 Iab star classified by Sota et al 2014, and 3 candidate O stars from \cite{Wramdemark1976}. If the 108 stars found conform to a \cite{kroupa2001} IMF, and a main-sequence mass of 7~M$_{\odot}$ may be associated with an effective temperature of 20000~K, the implied mass of stars in the association, is at least $\sim8\times10^3$ M$_{\odot}$.  On the same basis, the predicted number of O stars ($>15$~M$_{\odot}$) is 39 -- to be compared with our estimate of 37.  

\subsection{The changing on-sky OB star distribution with increasing extinction}
\label{sec:diffext}

\begin{figure*}
\centering
\includegraphics[height=0.95\textheight]{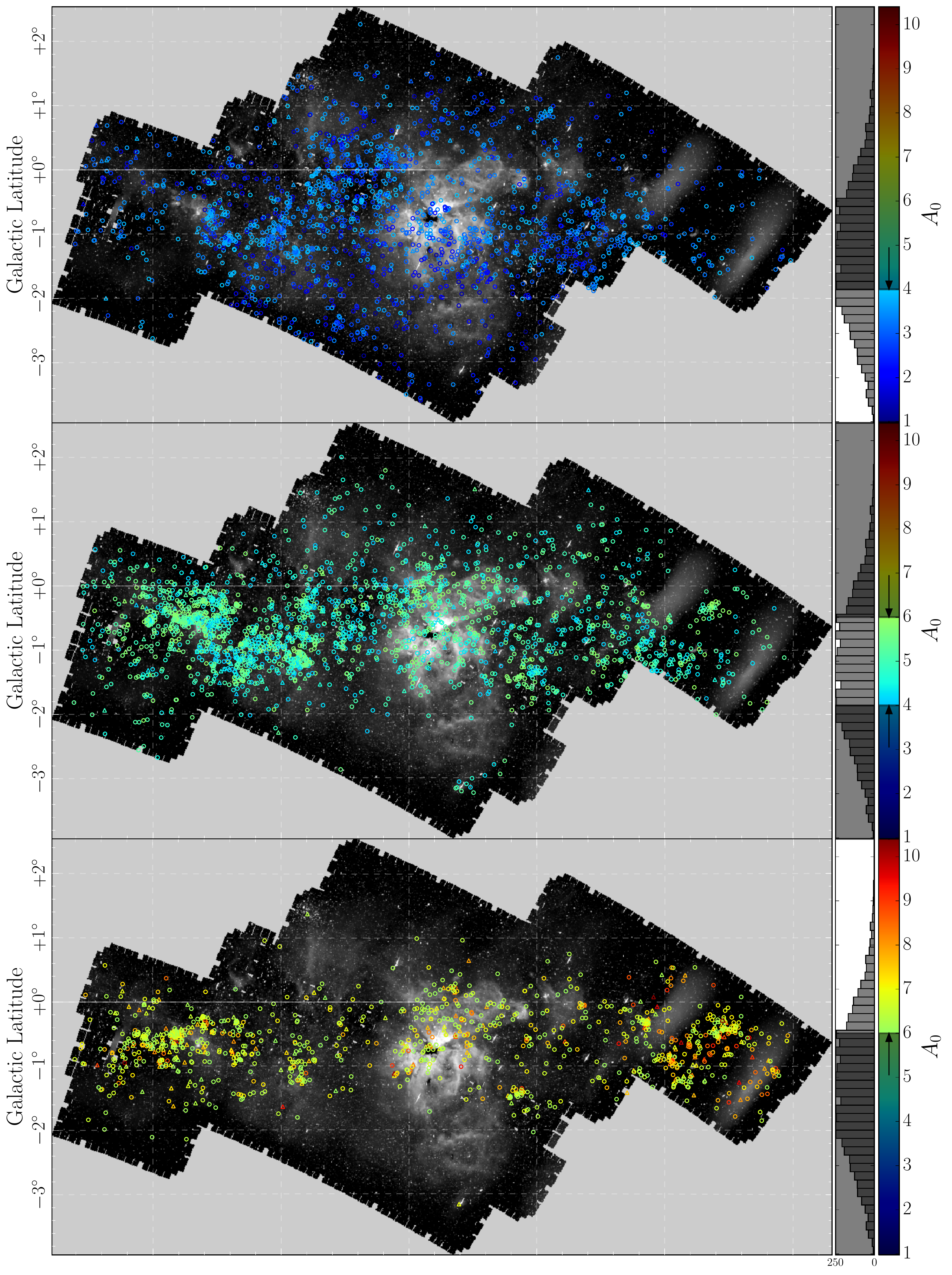}
\caption{Three plots showing the positions of OB stars through windows increasing in $A_0$ (from top to bottom). O stars and early B stars are represented as triangles and circles respectively. The histogram to the right of each on-sky map shows the full distribution in $A_0$, highlighting the range mapped to the left. The online version of this figure is animated, showing the changing OB star distribution as $A_0$ rises.}
\label{fig:extinctionmapOB}
\end{figure*}

\begin{figure*}
\centering
\includegraphics[width=0.65\textwidth]{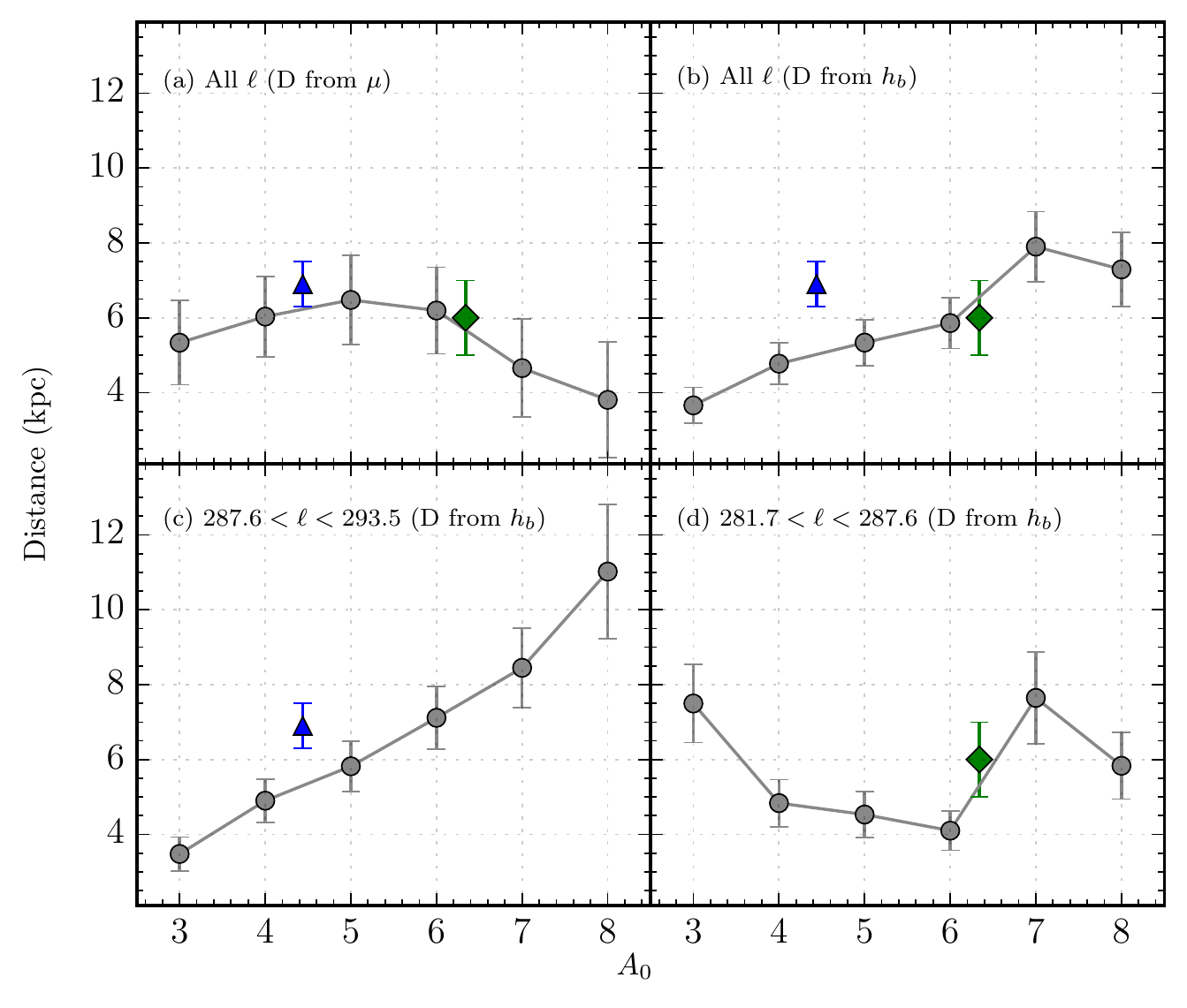}
\caption[Measuring extinction as a function of distance]{Panel (a): The mean distance and standard deviation obtained from the best-fitting distance modulus $\mu$ as a function of best-fit $A_0$.  Panel (b): here the distance is inferred as a function of $A_0$ based on the measured angular scale-height ($h_b$) of the newly uncovered OB star population, as described in the text.  This gives a much more orderly trend.  The blue triangle places NGC 3603 on the diagram according to the literature consensus on its reddening and distance, while the green diamond does the same for Westerlund 2.  Panels (c) and (d) are the same as (b) except that the longitude range has been split in half, with 3432 stars in the higher longitude (NGC 3603) half of the studied range in (c), and 1923 stars in the lower longitude range including Westerlund 2.}
\label{fig:d_a0}
\end{figure*}

Fig. \ref{fig:extinctionmapOB} shows the selected and accepted OB star candidates divided up between three different extinction windows ($A_0 \leq 4$, $4 < A_0 < 6$ and $A_0 \geq 6$). What stands out the most in this figure is that the distribution in Galactic latitude narrows as $A_0$ increases, revealing the receding Galactic thin disk.  To express this more precisely, the on-sky angular projection of the scale height of the OB stars exhibits clear shrinkage with increasing reddening -- hinting that reddening is statistically correlated with distance. We can use this effect to estimate the approximate distance ranges we are sampling as a function of rising extinction. This method turns out to be more informative than relying on the distance moduli derived from the SED fits, in which the strong correlation between $\mu$ and effective temperature leads to considerable imprecision.  Panel (a) of Fig. \ref{fig:d_a0} shows that the trend in $A_0$ versus distance, $D$ (from $\mu$) eventually loses credibility as distance would seem to decrease with extinction beyond $A_0 = 5$. In, view of this, we explore the use of the angular spread of the candidates as a proxy for distance.  This depends on the finding that the OB star scale height in the Galactic disk appears to be broadly independent of Galactocentric radius \citep{Paladini2004}.

\begin{figure*}
\centering
\includegraphics[width=0.75\textwidth]{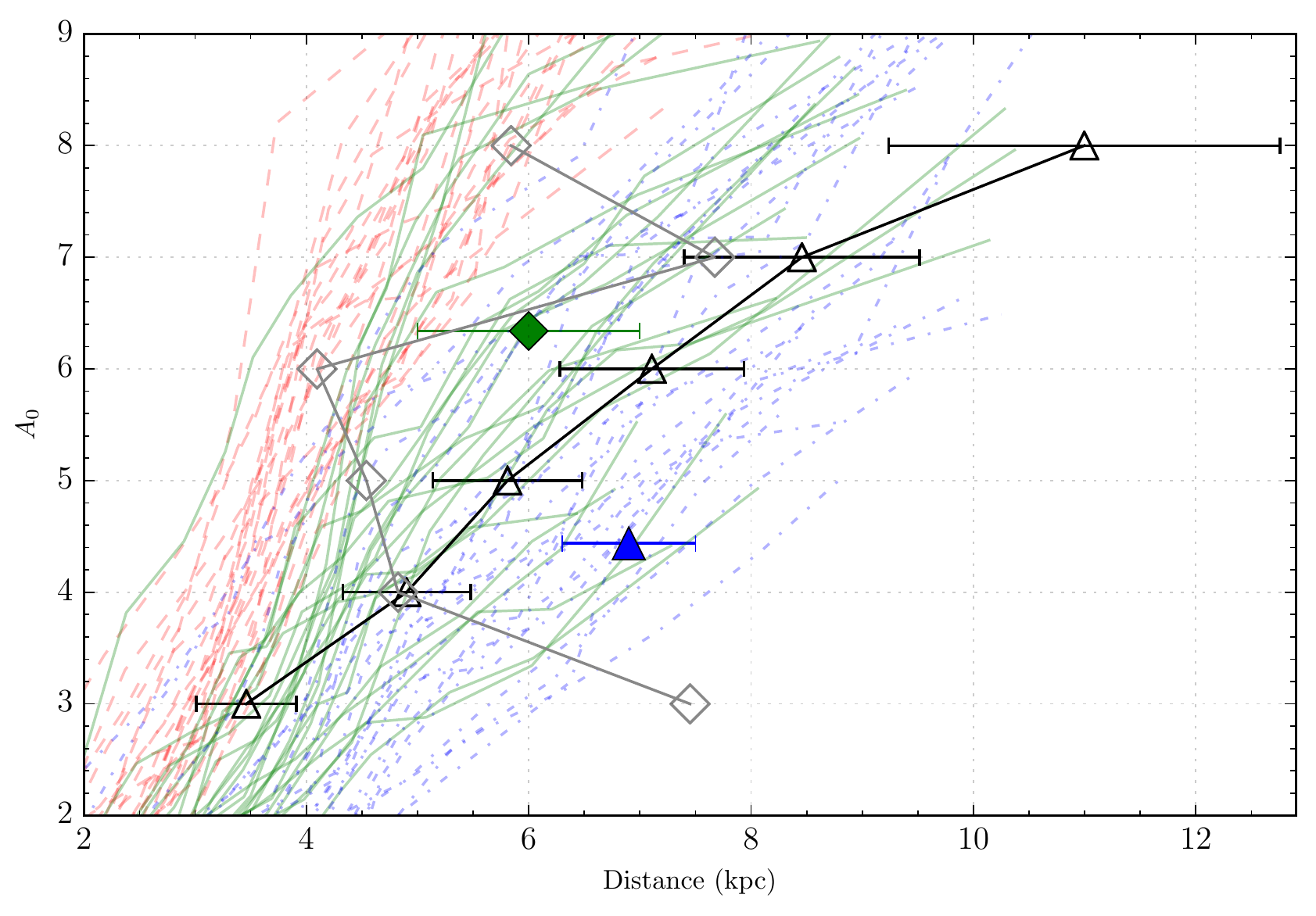}
\caption[Comparison of the derived values of extinction as a function of distance between \protect\cite{Marshalletal2006} and this study for sight lines sampling the Galactic longitude range.]{Comparison of the derived values of extinction as a function of distance between \protect\cite{Marshalletal2006} and this study. The diamond markers show the end of the footprint closer to the Carina Arm tangent, while the triangle markers show the opposite end around NGC 3603 (optical selections as in Fig.~\ref{fig:d_a0}). Error bars are included for the more credible longer-longitude sample (panel c in Fig.~\ref{fig:d_a0}). The red dashed lines and the blue dot-dashed lines are selections from \protect\cite{Marshalletal2006}, each sampling a square degree around respectively $\ell = 283^{\circ}$, $b = -0.5^{\circ}$ and $\ell = 291^{\circ}$, $b = -0.5^{\circ}$, at opposite ends of the footprint studied in this paper. The solid lines shown in green correspond to a square degree around $\ell = 286$, $b = -0.5^{\circ}$, near Wd 2. The blue triangle and green diamond show the corresponding values for NGC 3603 and Wd 2 respectively.}
\label{fig:marshall}
\end{figure*}

We bin the objects in extinction from $2.5 > 8.5$ in steps of $\Delta A_0 = 1$, anticipating that the OB stellar density, expressed as a function of Galactic latitude in each extinction slice, conforms to an isothermal population characterised by a single angular scale-height $h_b = h_{OB}/D$.  We adopt $h_{OB} = 45 \pm 5$~pc \citep{reed2000, bobylevetal2016}, and $D$ is to be regarded as the representative, if fuzzy, heliocentric distance to the reddening slice.  The scale height of OB stars, $h_{OB}$, is conventionally defined with reference to either a back-to-back exponential or to the square of a hyperbolic secant.  For our fitting exercise, we prefer the latter for mathematical convenience.  In this, we follow \citet{contivacca1990}.  Specifically, we apply to each reddening slice a fit of the form:  

\begin{equation}
N(b) = A~sech^2(\frac{b - b_0}{\sqrt{2}h_b})
\end{equation}

in which $A$ is a constant, proportional to the sample size, and $b_0$ is the Galactic latitude at which the best-fitting function peaks.  With $h_b$ the angular scale-height, this makes 3 fit parameters.  Panels (b), (c) and (d) of Fig. \ref{fig:d_a0} show the derived distances, $D$, per reddening slice, for three different Galactic longitude ranges. Panel (b) shows the entire longitude range of the studied area while panels (c) and (d) split the longitude range into two halves. Panel (c), where $287.6 < \ell < 293.2$, includes the part of the footprint that contains the cluster NGC 3603, while panel (d), where $282.0 < \ell < 287.6$ takes in the rest of the foot print containing Wd 2.  For use later, we set out the numerical results of panel (c) as Table~\ref{tbl:warp}.  The relevant panels in Fig.~\ref{fig:d_a0} include the distance and extinction to NGC 3603 (blue triangle) taken from \cite{sungandbessell2004} and the distance and extinction to Wd 2 (green diamond) taken from \cite{Dame2007} and \cite{mohr-smith2015} respectively. 

Panel (c) shows that for $287.6 < l < 293.2$ we have a well-behaved steady increase in estimated distance as a function of extinction of 0.8 mag/kpc.  This is the range within which the dependence of $A_0$ on Galactic longitude is also more organised, tending to rise as longitude increases (see Fig~\ref{fig:a0rv}.  The literature values of extinction and distance of NGC 3603 fit as expected into this trend, in showing it is located in a relative reddening hole \citep[see e.g.][]{pangetal2011}.

However, for $282.0 < l < 287.6$ shown in panel (d), there is no orderly correlation between extinction and inferred distance. This difference can be attributed to the fact that for the lower-longitude sight lines, the Carina Arm tangent \citep[$l \simeq 281$, see e.g.][]{Dame2007} is not far away, permitting a more rapid and chaotic accumulation of dust column density with increasing distance, looking almost directly along the Carina Arm.  In addition, our more limited latitude coverage at these longitudes may be implicated in the contradictory association of a long sightline with the lowest extinction bin (see Fig.~\ref{fig:extinctionmapOB}).
The lower overall density of candidate objects nearer the tangent direction may also be a clue that a more rapid accumulation of dust column brings the accessible sightlines more often than not to an end within the Arm.  In effect, the view rarely reaches much beyond Westerlund 2 at $D \sim 6$~kpc \citep{Dame2007}.  This stands in contrast to the general vicinity of NGC 3603 \citep[understood to be $\sim$7~kpc distant][]{sungandbessell2004}, where it appears from  panel (c) in Fig.~\ref{fig:d_a0} that OB stars are being picked up at heliocentric distances up to $\sim$ 10~kpc.

Our optically-based mapping can be compared with the $K$-band mapping carried out by \cite{Marshalletal2006}: produced a three dimensional extinction map of the Milky Way by comparing empirical colours of giant stars in 2MASS with the expected intrinsic colours from the Besan\c{c}on stellar population synthesis model. The angular resolution of this IR map is 0.25$^\circ$.  Fig.~\ref{fig:marshall} shows the derived extinction from \cite{Marshalletal2006} for three square-degree patches falling within the region under examination here. The sight lines shown in blue correspond to the higher Galactic longitude side of the region containing NGC 3603 ($290.5^\circ<\ell<291.5^\circ$ and $-1^\circ<b<0^\circ$) while those shown in red fall closest to the Carina tangent direction ($282.5^\circ<\ell<283.5^\circ$ and $-1^\circ<b<0^\circ$). The lines shown in green pick out a region in between ($285.5^\circ<\ell<286.5^\circ$ and $-1^\circ<b<0^\circ$).  The conversion factor between $A_K$ and $A_0$ is somewhat dependent on $R_V$: for the present purpose we will use 8, appropriate for $R_V\sim3.8$, to scale $A_K$ up to $A_0$.

We over-plot on the converted \cite{Marshalletal2006} relations in Fig.~\ref{fig:marshall} the extinction-distance trends already presented in panels (c) and (d) of Fig.~\ref{fig:d_a0}. The diamond markers represent the lower-longitude end of the footprint while the triangle markers show the higher-longitude end. Here we can see that the pattern apparent in the \cite{Marshalletal2006} data at $\ell \sim 291^{\circ}$ is consistent with our data (the blue curves, and triangles). On the other hand, the chaos in the OB-star data at lower longitudes (nearer the tangent, diamond symbols in the diagram), relative to the IR-based green and red curves, is underlined.  In the remaining discussion we restrict attention to the better behaviour seen at $\ell > 287.6^{\circ}$.

We now evaluate how the OB-star defined mid-plane, as specified by the fit parameter, $b_0$, deviates from the geometrically-flat mid-plane, at latitude $b_m$, along the observed pencil beam. The fit results relevant to this are collected together in Table~\ref{tbl:warp}.  The parameter, $b_m$, is itself subject to some uncertainty as it depends both on $D$ and on the assumed height of the Sun above the ideal flat mid-plane.  To complete our calculation we adopt estimates for $b_m$ that follow from the prescription set out by \cite{goodmanetal2014}, in which the Sun is offset above the mid-plane by 25 pc. The picture emerging from this is of a slowly growing offset, such that at $\sim$7~kpc the OB-star layer is about 60 pc below the geometric mid-plane, falling further to $80 - 100$~pc below at heliocentric distances of $9 - 10$~kpc.  At $\ell \sim 290^{\circ}$, this heliocentric distance range corresponds to a Galactocentric distance of $\sim10$ kpc.  The uncertainties in the estimates of warp in Table~\ref{tbl:warp}, arise only from the computed fit errors in distance-extinction trend shown in Fig.~\ref{fig:d_a0}: these are likely to be optimistic.  In particular, the estimated distance scale carries an as yet unknown bias due to the potentially simplifying assumption - implicit in the fit procedure - that the average rise in $A_0$ with distance has no strong latitude dependence near the mid-plane.

Neutral hydrogen 21-cm line observations of the Galactic gas disk imply little warping of the Galactic plane out to Galactocentric radii of $R_G \sim 9$ kpc \citep{kalberlaandkerp2009}. What we see here -- evidence of just the beginnings of a negative slowly growing warp -- fits in with this. Wellbeyond $R_G = 9$~kpc, they and \cite{levineetal2006} propose a rapid increase in the warp such that an offset of $\sim500$pc is reached at a Galactocentric distance of $R_G \sim 15$ kpc.

\begin{table*}
\caption{Data and results for the fits to extinction slices constructed from thecleaned OB star sample at Galactic longitudes $> 287.6^{\circ}$, using the functional form given in equation 9.  Here $b_m$, in the penultimate column is the true Galactic latitude of the geometrically flat mid-plane, according to \protect\cite{goodmanetal2014}.  In the last column, $Z_0$(OB) is the offset in parsec of the OB-star mid-plane from flat -- the estimated warp in other words.\label{tbl:warp}}
\begin{tabular}{ccccccc}

\hline
$A_0 [mag]$ & N & $b_0 [^{\circ}]$ & $h_b[^{\circ}]$ & D [kpc] & $b_m[^{\circ}]$ & $Z_0$ (OB) [pc] \\

\hline 
2.5--3.5 & 477 & $-0.60\pm0.06$ & $0.75\pm0.05$ & $3.5\pm0.5$ & -0.38 & $-13\pm-11$ \\
3.5--4.5 & 869 & $-0.66\pm0.03$ & $0.53\pm0.02$ & $4.9\pm0.6$ & -0.26 & $-35\pm-5$ \\
4.5--5.5 & 845 & $-0.66\pm0.02$ & $0.44\pm0.01$ & $5.8\pm0.7$ & -0.21 & $-46\pm-6$ \\
5.5--6.5 & 687 & $-0.65\pm0.02$ & $0.36\pm0.01$ & $7.1\pm0.8$ & -0.17 & $-60\pm-8$ \\
6.5--7.5 & 233 & $-0.63\pm0.03$ & $0.30\pm0.02$ & $8.5\pm1.1$ & -0.14 & $-72\pm-11$ \\
7.5--8.5 & 71 & $-0.66\pm0.04$ & $0.23\pm0.03$ & $11.0\pm1.8$ & -0.10 & $-108\pm-23$ \\
\hline
\end{tabular}
\end{table*}

\section{Discussion and conclusions}


In this paper, we applied our a method for the uniform and high-efficiency selection of OB stars and parametrization of their extinction, that was first laid out and validated by \cite{mohr-smith2015}. The selection here, using VPHAS+ and 2MASS survey data, has resulted in a catalogue of 14900 stars of which 5915 are high confidence O -- B2 stars, drawn from 42 square degrees in the Carina Arm region. The probable O stars number 905. The region charted here spans $\sim11$ degrees in Galactic longitude running from $\ell \sim 282^{\circ}$, close to the Carina Arm tangent direction, to $\ell \sim 293^{\circ}$, beyond the much studied massive cluster NGC 3603. By reaching down to $g= 20$\,mag roughly an order of magnitude increase is achieved relative to the numbers of known and candidate O -- B2 stars brighter than $\sim$12th magnitude: \cite{Reed2003} lists 845 confirmed/probable O and early-B stars in the same sky area, while the spectroscopic catalogue due to \cite{gosc2016} lists 106 O stars. Previous catalogues have included stars in the crowded hearts of NGC 3603 and Westerlund 2 that are missed here due to image blending. For example, \cite{melena2008} reported 51 OB stars in the brilliant, and packed, $\sim10$ arcsec core of NGC 3603 -- our catalogue lists 17 of them. This illustrates the complementarity that exists between focused studies and wide field survey selection.

\begin{figure*}
\centering
\includegraphics[width=0.95\textwidth]{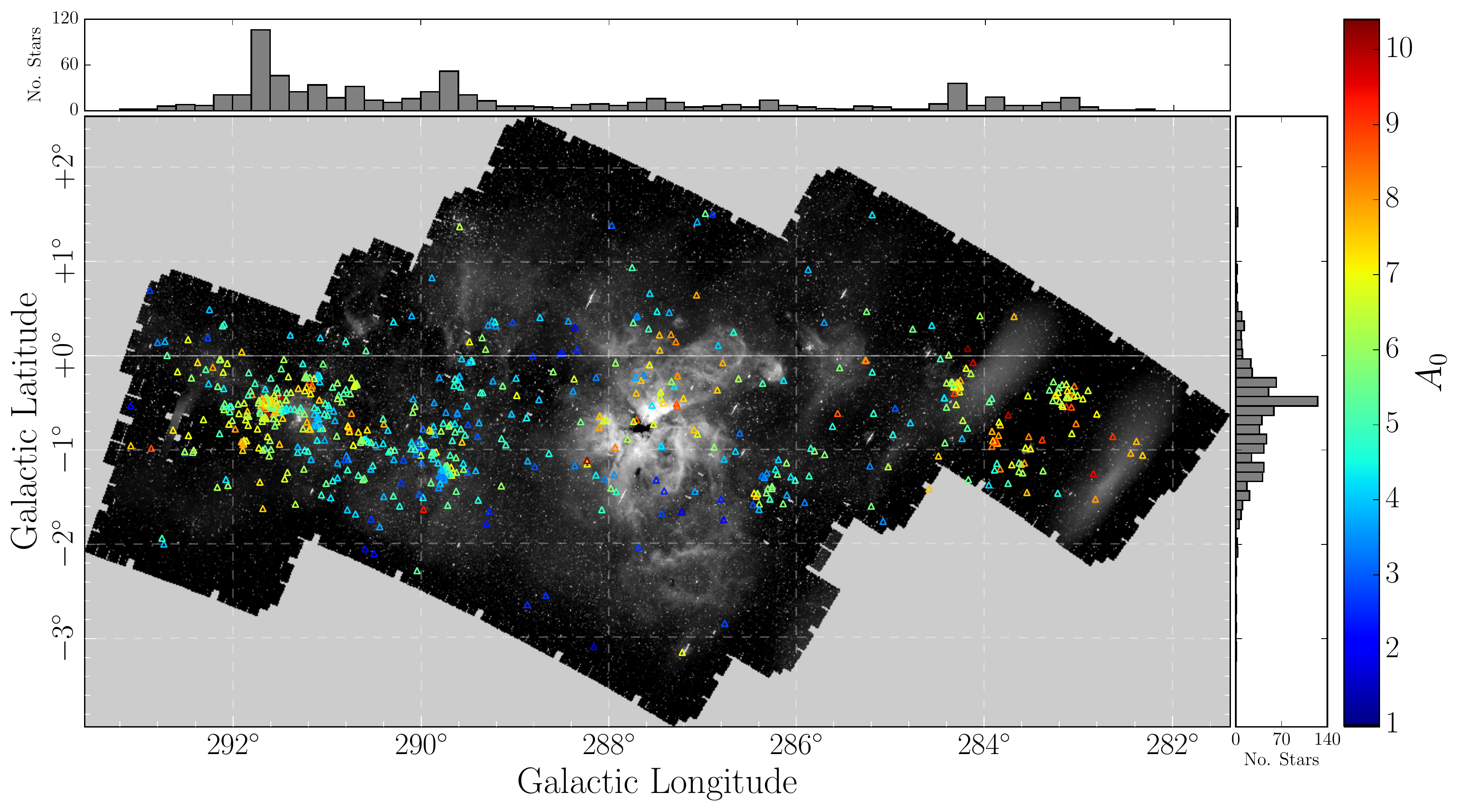}
\caption{Spatial distribution of candidate O stars across the Carina region, coloured by the extinction. The histograms above and to the right of the plot show the Galactic longitude and latitude distribution respectively}
\label{fig:extinctionmapO}
\end{figure*}

By bringing together VPHAS+ $u,g,r,i$ photometry with 2MASS $J,H,K$ photometry, we are able to determine both the value of the extinction, $A_0$, and the extinction law, as parametrised by $R_V$, to good precision: both are typically measured to better than 0.1 (magnitudes in the case of $A_0$).  Using a statistically much larger sample across a large area on the sky, we have confirmed and fleshed out previous results, reported by many authors, that sight lines into the Carina Arm require flatter reddening laws with $3.5\lesssim R_V \lesssim 4.0$ for $A_0 > 2.0$. 
Analysis in Section \ref{sec:diffext} has also demonstrated that extinction builds more quickly at Galactic longitudes closer to the Carina Arm tangent direction, where CO surface brightness is also known to rise to an evident peak \citep{Grabelsky1987}.  Indeed at $\ell \lesssim 286^{\circ}$, $g = 20$ is not yet deep enough to guarantee sightlines that pass through the Arm to much beyond a distance of 6--7~kpc.  However at greater longitudes, where the extinction builds at a mean rate of $\sim0.8$~mag kpc$^{-1}$, the distances sampled extend to beyond $\sim$10~kpc (or $R_G$ of at least 10 kpc also).

A crude impression of stellar effective temperatures (and hence distance moduli) is all that should be presumed to flow from the SED fits to the OnIR photometry of individual objects.  However, it is indeed possible to achieve a reasonable separation between candidate O and candidate early-B stars.  This was clearly confirmed in Section \ref{sec:spectra} where the analysis of AAOmega spectroscopy of a representative sample of 276 OB candidates was presented.  Among the better quality candidate objects ($\chi^2 < 7.82$) the cross-over rate, post spectroscopy, at the O/B boundary was only 10--20 \% (the higher percentage applying on the O star side of the boundary).  We also showed that the systematic offset between the photometric and spectroscopic effective temperature scales has only a modest effect on the derived extinction parameters for the B stars ($A_0\sim+0.09$ and $R_V\sim-0.07$) and an even less of an effect on the O stars ($A_0\sim+0.03$ and $R_Vs\sim-0.03$). This is a reflection of the convergence of the O star SED onto the Rayleigh-Jeans limit as effective temperature rises upwards of $\sim30000$~K. The follow-up spectroscopy demonstrated vey clearly that there are very low levels of contamination from late type stars within the selection (under 5\% of the accepted $\chi^2 < 7.82$ candidates). 

Further exploitation of the parameters derived from the photometry has also enabled an approximate distinction to be drawn between main sequence massive OB stars and lower luminosity hot stars (subdwarfs, post-AGB stars and others), and higher-luminosity evolved massive OB stars. Intriguingly, the one subluminous candidate to be followed up spectroscopically has emerged as a helium-enriched probable post-AGB object: the helium enrichment in particular is suggestive of material dredged up in a late helium flash  The narrow band H$\alpha$ filter in the VPHAS+ survey offers a clear path to the selection of emission line objects, and we find a fraction of B type emission line stars ($\sim7$\%) relative to the total B population that is consistent with the southern clusters study by \cite{mcswaingies2005}. The majority of emission line candidates are likely to be classical Be stars while a minority may also be examples of B[e] stars, luminous blue variables, or WR stars. The high-confidence broadband selection tends to disproportionately exclude new WR stars and many CBe stars, for the reason that these objects often present with NIR excesses that lead to inferior ($\chi^2 > 7.82$) SED fits.  

The catalogue created by this study is an opportunity to explore how a high angular resolution wide-field optical survey like VPHAS+ can bring a wider context to the study of star forming regions, away from the dense well-studied cores of open clusters and OB associations. The capability is now there to identify massive stars, wherever they are.
A compelling example of this is the very large number of candidate reddened O stars lying in an extended halo around the brilliant cluster, NGC 3603 at Galactic coordinates (291.6, -0.5).  This can be seen in Fig. \ref{fig:extinctionmapO} that shows the distribution of candidate O stars ($\ltf > 4.477$), on their own, across the whole region. Many objects are seen around NGC 3603, spread across a sky area about a degree across -- we find around 500 high-confidence O--B2 candidates in the extinction range $4.0 < A_0 < 7.0$, of which more than 100 are probable O stars.  Very recently, \cite{romanetal2016} have used near-infrared selection to identify 10 candidate early O stars in this same region and have confirmed them spectroscopically: all are in our catalogue with $\ltf$ and $A_0$ parameters from SED-fitting that are very close to their spectroscopic counterparts.  The shape on-sky of the O-star surface density enhancement fits in well with `star forming complex \#28', one of two G291 regions, identified via long-wavelength free-free emission by \cite{rahmanandmurray2010}.  Metaphorically speaking, NGC 3603 is the tip of a massive-star iceberg worthy of further investigation.

In Section 5.2 we also saw that at $\ell = 289.77^{\circ}$, $b = -1.22^{\circ}$, there is a clustering of OB stars in the vicinity of the bright O5 supergiant, LSS 2063, that before now had evaded optical detection and cataloguing.  This is sufficiently dispersed to be viewed as an OB association. As a further example, Fig.~\ref{fig:extinctionmapO} contains hints of another dispersed OB complex at $\ell \simeq 283^{\circ}$ and $b \simeq -0.5^{circ}$: in view of the similarly high extinctions involved ($A_0 \gtrsim 6$), it is likely to be about as distant as Westerlund 2. 

Many of the already well-known O stars in the Carina Nebula/Carina OB1 association at $l\sim288^{\circ}$ are missing from the selection presented here because they are brighter than our bright selection limit (at $V \sim 12$). However we do sweep up a handful of O star candidates and many B star candidates (see Fig. \ref{fig:extinctionmapOB}) which are far more extinguished than the typical values of $A_V\sim2.5$ of known clusters such as Trumpler 14 and 16 \citep{huretal2012}. These objects are likely to be situated behind or deeper into the nebula than the well studied cluster stars.  Indeed the closest distance for the majority of massive OB stars in the catalogue presented here is expected to be 2.5--3 kpc, the range in distance within which the Carina Nebula is likely to be located.

The full catalogue of blue selected objects, including spectroscopic data where they exist, is available in machine-readable form as supplementary material and will also become available via CDS: the Appendix to this paper provides a specification of the columns making up the catalogue. 

The ability to efficiently and reliably select and measure thousands of fainter OB stars opens the way to using these intrinsically luminous objects to probe the structure of the Galactic disk in a more fine-grained fashion than has been possible before.  Through sampling the field thoroughly, for the first time, as well as massive star clusters, this study has provided systematic evidence that the latter promote somewhat flatter optical extinction laws (i.e. increased $R_V$) alongside greater $A_0$.  The results of Fitzpatrick \& Massa (2007, Figure 11) hint at this but lack the sampling in the field.  The Carina Nebula itself has emerged as an especially striking example of raised $R_V$. Around NGC 3603, we obtain a median extinction of 6 magnitudes, and median $R_V \simeq 3.75$, to be compared with below 5 magnitudes and 3.55 respectively, recently measured inside the cluster \citep{pangetal2016}.  Further investigation of the subtle differentiations beginning to appear will help fill out the picture of dust processing arising from star formation. 

The Galactic latitude distribution of the OB stars has been shown to imply incipient disk warping along sightlines near $\ell = 290^{\circ}$: our approximate method of inferring distance via reddening suggests heliocentric distances on the order of 10~kpc are reached and that the OB-star layer offset below the mid-plane at this distance is likely to be 80--100~pc (Sec.~\ref{sec:diffext}).  Before now, similar efforts using OB stars to trace disk warping have had to work with very much lower sky densities of bright objects \citep[1300 stars over a third of the Galactic disc,][]{reed1996} or accept data inhomogeneity \citep[600 Carina region stars compiled from 4 catalogues][]{kaltcheva2010}.  These authors, and \cite{Graham1970}, proposed stronger warping, compared to the new result, in part because of over-estimation of distances driven by under-estimation of visual extinctions (through the adoption of $R_V \simeq 3.1$).  

The deep census provided here will be extended across the rest of the Galactic plane in future publications as VPHAS+, along with its northern counterparts, The INT/WFC Photometric H$\alpha$ Survey \citep[IPHAS][]{drewetal2005} and The UV Excess Survey \citep[UVEX][]{grootetal2009}, progress to fully-calibrated completion.  These are set to combine transformationally with forthcoming Gaia astrometry -- reaching to a similar faint limit -- in enabling much more precise analysis both of Galactic disk structure, and also of the ecology and life cycles of massive Galactic clusters and their environments. 

\section*{Acknowledgements}

This paper is based on data products from observations made with ESO Telescopes at the La Silla Paranal Observatory under programme ID 177.D-3023, as part of the VST Photometric Hα Survey of the Southern Galactic Plane and Bulge (VPHAS+, www.vphas.eu).  All VPHAS+ data are processed by the Cambridge Astronomical Survey Unit at the Institute of Astronomy in Cambridge: we would particularly like to thank Eduardo Gonzalez-Solares and Mike Irwin for their efforts over the years.
Use is also made of data products from the Two Micron All Sky Survey, which is a joint project of the University of Massachusetts and the Infrared Processing and Analysis Center/California Institute of Technology, funded by the National Aeronautics and Space Administration and the National Science Foundation. WISE and AAT service programme acknowledgements. We thank the anonymous referee of this paper for their supportive remarks.

This research made use of \textsc{Astropy}, a community-developed core Python package for Astronomy \citep{AstropyCollaboration2013} and \textsc{TopCat} \citep{Taylor2005}.

MM-S acknowledges a studentship funded by the Science and Technology Facilities Council (STFC) of the United Kingdom (ref. ST/K502029/1). JED, MM and GB acknowledge the support of research grants funded by the STFC (ref. ST/J001333/1 and ST/M001008/1). SS-D acknowledges funding by the Spanish Ministry of Economy and Competitiveness (MINECO) under the grants AYA2010-21697-C05-04, AYA2012-39364-C02-01, and Severo Ochoa SEV-2011-0187.  NJW acknowledges the support of a Research Fellowship awarded by the Royal Astronomical Society (to September 2015), and an STFC Ernest Rutherford Fellowship (from October 2016, ref ST/M005569/1).

\footnotesize{
\bibliographystyle{mn2e}
\bibliography{year1report}
}

\clearpage
\begin{appendix}

\section{Database description}

The catalogue of the complete set of 14900 candidate OB stars selected from the VPHAS+ $(u-g,g-r)$ colour-colour diagram is provided as a machine readable database in the supplementary material.  This will be made available via CDS.

The source positions and multi-band photometry on which the initial selection was made is provided in the first 17 columns. The RA and DEC reported are derived from the $r$ band observations obtained with the $u$ and $g$ filters. Columns 18--24 list the 2MASS cross-match magnitudes and errors: note that the database does not include blue-selected objects {\em without} 2MASS counterparts.  All photometric magnitudes are reported in the Vega system. For 152 ($\sim 1 \%$) of the selected objects, H$\alpha$ magnitudes are presently unavailable: where this applies the database entry is left blank. Similarly the second $r$ magnitude is missing for 59 stars.  Column 24 lists the VPHAS+/2MASS cross-match separation.

The next columns, 25--38, summarise the results of the SED fits, presenting the median of the posterior distribution for all four parameters along with errors derived from the 16$^{\rm th}$ and 84$^{\rm th}$ percentiles (essentially 1-$\sigma$ errors).  The final important piece of information from the fits is the value of $\chi^2$ computed for the best-fitting SED, given in column 37: where this value is $<7.82$ and $\ltf > 4.30$ (to two significant figures), column 38 in the database is set to true. We caution that the parameters for objects with $\chi^2 > 7.82$ are less reliable: these objects are more heterogeneous and will include emission line objects, some normal OB stars as well as contaminant objects. Columns 39--41 are similar boolean columns that identify candidate emission line stars (as determined from $(r - H\alpha)$ excess), and sub-luminous and evolved non-main-sequence OB candidates.

Where alternative names for an object already exist in the SIMBAD database, we write this into column 42, the  `Notes' column.  We also note any other specific characteristics of the object that have already come to attention: in particular, if an AAOmega spectrum exists that shows the object to have line emission at H$\beta$ we note it here.  The small number of known contaminant late-type stars are noted here also.

The final set of columns (43--53) records whether an AAAOmega spectrum exists and reports numerical quantities derived from model atmosphere fits as appropriate.  For the contaminant stars, all numerical quantities are omitted. 

Table \ref{tbl:photom_params} below provides individual specifications for all the columns in the database.

\clearpage
\onecolumn
\captionof{table}{Specifications for all the columns in the database provided in the supplementary materials. \label{tbl:photom_params}} 

\medskip

\begin{center}
\begin{tabular}{lllll}

\hline \noalign{\smallskip}
\bf Column &\bf Name & \bf Description & \bf Units & \bf Type \\ \noalign{\vspace{0.1cm}}
\hline \noalign{\smallskip}

1 & Name & `VPHAS\_OB1\_NNNNN', where `NNNNN' is ordered by Galactic longitude & & String \\ \noalign{\vspace{0.05cm}}
2 & RA & Right Ascension (J2000) & Deg. & Float  \\ \noalign{\vspace{0.05cm}}
3 & DEC & Declination (J2000) & Deg. & Float \\ \noalign{\vspace{0.05cm}}
4 & GAL\_LONG & IAU 1958 Galactic Longitude & Deg. & Float \\ \noalign{\vspace{0.05cm}}
5 & GAL\_LAT & IAU 1958 Galactic Latitude & Deg. & Float \\ \noalign{\vspace{0.05cm}}
6 & u & VPHAS+ u band & Vega mag & Float \\ \noalign{\vspace{0.05cm}}
7 & u\_err & VPHAS+ u band photometric uncertainty & Vega mag & Float \\ \noalign{\vspace{0.05cm}}

8 & g & VPHAS+ g band & Vega mag & Float \\ \noalign{\vspace{0.05cm}}
9 & g\_err & VPHAS+ u band photometric uncertainty & Vega mag & Float \\ \noalign{\vspace{0.05cm}}
10 & r & VPHAS+ r band taken with $u$ and $g$ (used in SED fits) & Vega mag & Float \\ \noalign{\vspace{0.05cm}}
11 & r\_err & VPHAS+ r band photometric uncertainty & Vega mag & Float \\ \noalign{\vspace{0.05cm}}
12 & r2 & VPHAS+ second r band, taken with $i$ and H$\alpha$ & Vega mag & Float \\ \noalign{\vspace{0.05cm}}
13 & r2\_err & VPHAS+ second r band photometric uncertainty & Vega mag & Float \\ \noalign{\vspace{0.05cm}}
14 & i & VPHAS+ i band & Vega mag & Float \\ \noalign{\vspace{0.05cm}}
15 & i\_err & VPHAS+ i band photometric uncertainty & Vega mag & Float \\ \noalign{\vspace{0.05cm}}
16 & Ha & VPHAS+ H$\alpha$ band & Vega mag & Float \\ \noalign{\vspace{0.05cm}}
17 & Ha\_err & VPHAS+ H$\alpha$ band photometric uncertainty & Vega mag & Float \\ \noalign{\vspace{0.05cm}}
18 & J & 2MASS J band & Vega mag & Float \\ \noalign{\vspace{0.05cm}}
19 & J\_err & 2MASS J band photometric uncertainty & Vega mag & Float \\ \noalign{\vspace{0.05cm}}
20 & H & 2MASS H band & Vega mag & Float \\ \noalign{\vspace{0.05cm}}
21 & H\_err & 2MASS H band photometric uncertainty & Vega mag & Float \\ \noalign{\vspace{0.05cm}}
22 & K & 2MASS K band & Vega mag & Float \\ \noalign{\vspace{0.05cm}}
23 & K\_err & 2MASS K band photometric uncertainty & Vega mag & Float \\ \noalign{\vspace{0.05cm}}
24 & xmatch\_r & The cross match distance between VPHAS+ and 2MASS	& arcseconds & Float \\ \noalign{\vspace{0.05cm}}

25 & logTeff & Estimated effective temperature from photometric fits (median of posterior) & log(K) & Float \\ \noalign{\vspace{0.05cm}}
26 & logTeff\_eu & Upper uncertainty on logTeff (84$^{\rm th}$ percentile of posterior) & log(K) & Float \\ \noalign{\vspace{0.05cm}}
27 & logTeff\_el & Lower uncertainty on logTeff (16$^{\rm th}$ percentile of posterior) & log(K) & Float \\ \noalign{\vspace{0.05cm}}

28 & A0 & Estimated extinction, $A_0$, from photometric fits (median of posterior) & mag & Float \\ \noalign{\vspace{0.05cm}}
29 & A0\_eu & Upper uncertainty on A0 (84$^{\rm th}$ percentile of posterior) & mag & Float \\ \noalign{\vspace{0.05cm}}
30 & A0\_el & Lower uncertainty on A0 (16$^{\rm th}$ percentile of posterior) & mag & Float \\ \noalign{\vspace{0.05cm}}

31 & Rv & Estimated extinction law parameter, $R_V$, from photometric fits (median of posterior) & & Float\\ \noalign{\vspace{0.05cm}}
32 & Rv\_eu & Upper uncertainty on RV (84$^{\rm th}$ percentile of posterior) &   & Float \\ \noalign{\vspace{0.05cm}}
33 & Rv\_el & Lower uncertainty on RV (16$^{\rm th}$ percentile of posterior) &   & Float \\ \noalign{\vspace{0.05cm}}

34 & mu & Estimated distance modulus, $\mu$, from photometric fits (median of posterior) & mag & Float \\ \noalign{\vspace{0.05cm}}
35 & mu\_eu & Upper uncertainty on mu (84$^{\rm th}$ percentile of posterior) & mag & Float \\ \noalign{\vspace{0.05cm}}
36 & mu\_el & Lower uncertainty on mu (16$^{\rm th}$ percentile of posterior) & mag & Float \\ \noalign{\vspace{0.05cm}}

37 & chi2 & $\chi^2$ value for photometric fit &  & Float  \\ \noalign{\vspace{0.05cm}}

38 & goodOB & Indicates if object has $\chi^2 < 7.82$ and $\log(\rm T_{eff}) > 4.30$, rounded to significant figures & & Boolean \\ \noalign{\vspace{0.05cm}}

39 & EM & Indicates if star is selected as an emission line object & & Boolean \\ \noalign{\vspace{0.05cm}}

40 & SUB & Indicates if star is selected as a sub-luminous object & & Boolean \\ \noalign{\vspace{0.05cm}}

41 & LUM & Indicates if star is selected as an over-luminous object & & Boolean \\ \noalign{\vspace{0.05cm}}

42 & Notes & Comments, including SIMBAD cross identification name where it exists &  & String \\ \noalign{\vspace{0.05cm}}

43 & SPEC & Indicates if star has an AAOmega spectrum & & Boolean \\ \noalign{\vspace{0.05cm}}
44 & spec\_mod & The model grid used for fitting: TLUSTY or FASTWIND & & String \\  \noalign{\vspace{0.05cm}}
45 & speclogTeff & Estimated effective temperature derived from spectroscopic fit (median of posterior) & log(K) & Float \\ \noalign{\vspace{0.05cm}}
46 & speclogTeff\_eu & Upper uncertainty on specTeff (84$^{\rm th}$ percentile of posterior) & log(K) & Float \\ \noalign{\vspace{0.05cm}}
47 & speclogTeff\_el & Lower uncertainty on specTeff & log(K) & Float \\ \noalign{\vspace{0.05cm}}

48 & logg & Estimated surface gravity derived from spectroscopic fits (median of posterior) & dex & Float \\ \noalign{\vspace{0.05cm}}
49 & logg\_eu & Upper uncertainty on logg (84$^{\rm th}$ percentile of posterior) & dex & Float \\ \noalign{\vspace{0.05cm}}
50 & logg\_el & Lower uncertainty on logg & dex & Float \\ \noalign{\vspace{0.05cm}}

51 & vsinsi & Estimated rotational velocity derived from spectroscopic fits (median of posterior) & kms$^{-1}$ & Float \\ \noalign{\vspace{0.05cm}}
52 & vsini\_eu & Upper uncertainty on vsini (84$^{\rm th}$ percentile of posterior) & kms$^{-1}$ & Float \\ \noalign{\vspace{0.05cm}}
53 & vsini\_el & Lower uncertainty on vsini & kms$^{-1}$  & Float\\ \noalign{\vspace{0.05cm}}

\hline
\hline

\end{tabular}
\end{center}
\end{appendix}

\end{document}